\documentclass[twocolumn]{aastex62}

\usepackage{longtable}
\usepackage{graphicx}	
\usepackage{amsmath}	
\usepackage{amssymb}	
\usepackage{color}

\usepackage[figuresright]{rotating}
\usepackage{threeparttable}
\usepackage{enumerate}
\usepackage{verbatim}
\usepackage{url}
\usepackage{float}
\usepackage{rotating}
\usepackage[T1]{fontenc}
\usepackage{hyperref}

\received{xxx}
\revised{xxx}
\accepted{xxx}
\submitjournal{ApJ}

\shorttitle{Results of 12 Years of Pulsar Timing at Nanshan - I}
\shortauthors{S.J.Dang et al.}

\begin{document}

\title{Results of 12 Years of Pulsar Timing at Nanshan - I}

\correspondingauthor{J. P. Yuan}
\email{yuanjp@xao.ac.cn}
\correspondingauthor{N. Wang}
\email{na.wang@xao.ac.cn}

\author[0000-0002-0786-7307]{S.J.Dang}
\affil{XinJiang Astronomical Observatory, CAS, Urumqi, XinJiang 830011, China\\}
\affiliation{University of Chinese Academy of Sciences,19A Yuquan Road, BeiJing 100049,China\\}

\author{J.P.Yuan}
\affiliation{XinJiang Astronomical Observatory, CAS, Urumqi, XinJiang 830011, China\\}
\affiliation{Key Laboratory of Radio Astronomy, Chinese Academy of Sciences, Nanjing 210008, China\\}

\author{R. N. Manchester}
\affiliation{CSIRO Astronomy and Space Science, Australia Telescope National Facility, PO Box 76, Epping, NSW 1710, Australia\\}

\author{L.Li}
\affiliation{XinJiang Astronomical Observatory, CAS, Urumqi, XinJiang 830011, China\\}
\affiliation{School of Physics Science and Technology, Xinjiang University, Urumqi, Xinjiang}

\author{N.Wang}
\affiliation{XinJiang Astronomical Observatory, CAS, Urumqi, XinJiang 830011, China\\}
\affiliation{Key Laboratory of Radio Astronomy, Chinese Academy of Sciences, Nanjing 210008, China\\}

\author{J.B.Wang}
\affiliation{XinJiang Astronomical Observatory, CAS, Urumqi, XinJiang 830011, China\\}
\affiliation{Key Laboratory of Radio Astronomy, Chinese Academy of Sciences, Nanjing 210008, China\\}

\author{G. Hobbs}
\affiliation{CSIRO Astronomy and Space Science, Australia Telescope National Facility, PO Box 76, Epping, NSW 1710, Australia\\}

\author{Z.Y.Liu}
\affiliation{XinJiang Astronomical Observatory, CAS, Urumqi, XinJiang 830011, China\\}
\affiliation{Key Laboratory of Radio Astronomy, Chinese Academy of Sciences, Nanjing 210008, China\\}

\author{F.F.Kou}
\affiliation{XinJiang Astronomical Observatory, CAS, Urumqi, XinJiang 830011, China\\}
\affiliation{CAS Key Laboratory of FAST, National Astronomical ObservatoriesChinese Academy of Sciences,Beijing, China\\}

\begin{abstract}
We have used the Nanshan 25-m Radio Telescope at Xinjiang Astronomical Observatory to obtain timing observations of 87 pulsars from 2002 July to 2014 March. Using the "Cholesky" timing analysis method we have determined positions and proper motions for 48 pulsars, 24 of which are improved positions compared to previously published values. We also present the first published proper motions for nine pulsars and improved proper motions for 21 pulsars using pulsar timing and position comparison method. The pulsar rotation parameters are derived and are more accurate than previously published values for 36 pulsars. Glitches are detected in three pulsars: PSRs J1722$-$3632, J1852$-$0635 and J1957+2831. For the first two, the glitches are large,  with $\Delta\nu_g/\nu > 10^{-6}$, and they are the first detected glitches in these pulsars. PSR J1722$-$3632 is the second oldest pulsar with large glitch. For the middle-age pulsars ($\tau_c > 10^5$~yr), the calculated braking indices, $|n|$, are strongly correlated with $\tau_c$ and the numbers of positive and negative values of $n$ are almost equal. For young pulsars ($\tau_c < 10^5$~yr), there is no correlation between $|n|$ and $\tau_c$ and most have $n>0$.
\end{abstract}


\section{Introduction}
Pulsar timing is employed to determine pulsar positions and proper motions \citep{hlk+04,zhw+05,lwy+16}, test the general theory of relativity \citep[e.g.,][]{ht75,ksm+06}, detect gravitational waves \citep[e.g.,][]{hd83,pddk+19}. The pulsar timing technique \citep{MTJ77,BH86,LK04} allows observed topocentric pulse times of arrival (ToAs) to be compared with a model which contains information about the pulsar astrometric, orbital and rotational parameters. The differences between the actual ToAs and the predicted ToAs are known as pulsar timing residuals.
For a perfect model, the timing residuals would be dominated by measurement uncertainties. Any other features observed in the timing residuals indicate the presence of un-modelled effects which may include calibration uncertainties, spin-down irregularities or the timing signal caused by gravitational waves. The spin-down of pulsars is usually remarkably stable and predictable. However, timing observation of pulsars with low characteristic age have revealed two types of rotation irregularities, namely, glitches and timing noise.

Glitches are manifested as a sudden, discontinuous increase of the rotation frequency, in some cases followed by a period of relaxation \citep{ymw+10, ywm+10, els+11, ymh+13}. About 550 glitches have been detected in pulsars\footnote{http://www.atnf.csiro.au/research/pulsar/psrcat/glitchTbl.html and http://www.jb.man.ac.uk/pulsar/glitches/gTable.html}.
Possible models for glitches include starquakes \citep{bpp+1969} and coupling between the interior superfluid and the inner crust, resulting in a rapid transfer of angular momentum to the outer crust of the star \citep{ai+75}.

Timing noise consists of unexpected, and thus un-modelled, time-correlated features in the timing residuals relative to the existing slowdown model. It is seen in most pulsars at some level. One possible cause for these correlated variations is intrinsic "spin noise" which results from rotational irregularities of the neutron star itself. This is generally regarded as a stochastic noise process with a red power spectrum, with studies of large samples being carried out in the past \citep[e.g.,][]{hlk+10}. Low-frequency structures observed in pulsar data sets have been modelled by, for example, random processes \citep{bgh+72}, superfluid turbulence \citep{ml+14} and variations in spin-down torque \citep{ulw+06,lhk+10}. However, the physics underlying timing noise is still largely unknown. A better understanding of pulsar timing noise may provide an insight to the interior structure of neutron stars \citep{ai+75, els+11, ymh+13}.

Pulse frequencies ($\nu$), frequency first time-derivatives ($\dot\nu$), frequency second time-derivatives ($\ddot\nu$), and pulsar positions and proper motions,  are important parameters for pulsars. The pulse frequency and its time derivative determine the pulsar characteristic age, $\tau_c \equiv -\nu/(2\dot\nu)$, an upper limit on the true age of the pulsar. If the $\nu$, $\dot\nu$ and $\ddot\nu$ are known, the braking index $n=\nu\ddot\nu/\dot\nu^2$ may be computed. Because of the timing irregularities, the pulsar rotation parameters, positions and proper motions will deviate from the predicted value. High-order polynomial fitting \citep{zhw+05} and harmonic whitening \citep{hlk+04} were used to reduce the effect of timing noise. \citet{chc+11} showed that these two methods often lead to severe underestimates of the parameter uncertainties and present the  "Cholesky" method based on a generalized-least-squares fitting procedure which can whiten the residuals to provide unbiased parameter measurements in the presence of timing noise. The method has been implemented as the \textsc{spectralModel} plug-in of the \textsc{Tempo2} software package \footnote{https://sourceforge.net/projects/tempo2/} \citep{hem+06} and has been applied to determine the positions and proper motions of millisecond pulsars \citep{rhc+16} and young pulsars \citep{lwy+16}.

In this paper, we use the "Cholesky" method to determine the position and proper motion for 48 pulsars and the rotation parameters for 87 pulsars. In Section 2, we introduce the pulsar observation system at the Nanshan 25-m radio telescope of the Xinjiang Astronomical Observatory (XAO) and the data processing method. The results of data analysis are given in Section 3. Section 4 presents the discussion and Section 5 presents the conclusions from our work.

\section{Observations and data analysis}
Timing observations of pulsars using the Nanshan 25-meter Radio Telescope of Xinjiang Astronomical Observatory, Chinese Academy of Sciences, started in 2000 January with a dual-channel room-temperature receiver. Since July 2002, a cryogenic receiver has been used for regular pulsar timing observations. The receiver bandwidth is 320~MHz, from 1380~MHz to 1700~MHz, with a center frequency of 1540~MHz. Prior to 2010, observations were made using an analogue filter bank (AFB) with $2\times128\times2.5$~MHz channels \citep{wmz+01}. A digital filter bank (DFB) constructed by the CSIRO Australia Telescope National Facility (ATNF) has been used since January 2010. The DFB is capable of processing up to 1~GHz of bandwidth with four primary modes of operation, i.e., folding, search, spectrometer and baseband outputs. Pulsar observing systems require a method of dividing the observed bandwidth into narrow channels to counter the effects of interstellar dispersion. The ATNF DFB systems are based on implementation of a polyphase filter in field programmable gate array (FPGA) processors \citep{mhb+13}. The DFB hardware consists of three main parts: a high speed analog-to-digital converter, the polyphase filterbank and a Pulsar Processing Unit (PPU). The PPU contains a correlator giving  polarisation products and then folds data at the pulsar period. The DFB was configured to give $1024\times0.5$~MHz channels with 8-bit sampling. For each observation, the integration time is 4 to 16 minutes, and the observing cadence is about one session every nine days. In this paper, the time span of the data for two pulsars is from January 2000 to March 2014 and for the others from July 2002 to March 2014.

After obtaining the data, we applied the \textsc{psrchive} software \citep{hvm+04} for offline processing, with the pulsar ephemeris from the ATNF Pulsar Database\footnote{http://www.atnf.csiro.au/research/pulsar/psrcat/}. Firstly, the \textsc{psrchive} software \citep{vdo12} was employed to remove radio frequency interference and to incoherently de-disperse the data, and then sum in frequency, time and polarization to produce a mean pulse profile. Secondly, we summed all of the observed profiles to produce a high signal-to-noise ratio profile and then used the \textsc{paas} software package to form a noise-free standard profile for each pulsar. Finally, we cross-correlated the mean pulse profile for each observation with the standard profile to obtain topocentric pulse ToAs. As the data were acquired simultaneously by the AFB and DFB from January 2010 to March 2014, the \textsc{splug} plug-in of \textsc{Tempo2} software package was employed to select the ToAs with the least uncertainty. To remove the effects of the Earth's motion, these ToAs must be converted to the solar-system barycenter (SSB), which can be regarded as the origin of an inertial reference system. The time system used was Barycentric Coordinate Time (TCB) and we used the solar-system ephemeris DE421 \citep{fwb+09}. The observatory clock is aligned with UTC via GPS and the clock offsets are corrected to refer the ToAs to International Atomic Time (TAI) by \textsc{Tempo2}. \textsc{Tempo2} then fits a pulsar timing model to the corrected ToAs to obtain a timing solution. An offset between the AFB and DFB ToAs was included in the timing model.

The timing analysis also employed the \textsc{spectralModel} plug-in of the \textsc{Tempo2} software package. The implementation steps are as follows:
\begin{enumerate}[(i)]
\item Using the \textsc{efacEquad} plugin to determine the scaling factor "EFAC" for the original uncertainties and the extra noise in quadrature  "EQUAD" to correct the measured  uncertainties.\footnote{Artefacts in the profiles from instrumental instabilities, imperfect profile templates, un-excised radio frequency interference (RFI) and intrinsic pulse jitter may lead to errors in the ToA uncertainty estimates. The measured ToA uncertainties can be multiplied by a scaling factor EFAC or a correction term EQUAD added quadratically to allow for these additional errors. Here, the new uncertainty, $\sigma^\prime$ is related to the original uncertainty $\sigma$ by ${\sigma^\prime}^2 = {\rm EFAC}^2 \times (\sigma^2 + {\rm EQUAD}^2)$.}
\item Estimate an initial model of the timing noise with the \textsc{spectralModel} plugin. The analytic model for the spectrum of red noise is as follows:
\begin{equation}
\label{eq:pf}
 P(f) = \frac{A}{[1+(f/f_c)^2)]^{\alpha/2}}.
\end{equation}
where A is an amplitude (actually the noise power at zero frequency), $\alpha$ is the spectra index and $f_c$ is the corner frequency below which the noise power spectrum plateaus. The corner frequency $f_c$ should not be $f_c  <  1/T_{obs}$ because fitting $\nu$ and $\dot\nu$ will flatten the spectrum below this frequency \citep{chc+11}. 
\item If the spectrum of the initial red noise shows a peak in the frequency of 1~yr$^{-1}$, we include a fit for position along with other pulsar parameters, including the initial red-noise model, using the "Cholesky" least-squares fitting procedure \citep{chc+11}.
\item Determine a better red-noise model utilizing the improved parameters (excluding fitting for the position and proper motion).
\item Employ the better red-noise model and the "Cholesky" least-squares fitting procedure to fit the pulsar parameters (including position and proper motion).
\item Improve the pulsar parameters and red-noise model by iterating previous steps until the results converge. Then save the final improved pulsar parameters (including position, proper motion and red-noise model).
\end{enumerate}

\section{Results}
\subsection{Positions and proper motions of 48 pulsars}\label{sec:Positions and proper motions}
New positions and proper motions have been determined from the Nanshan timing observations using the methods described above for 48 pulsars. Proper motions have also been obtained by comparing the timing positions with those from the literature  with earlier epochs \citep[cf.,][]{zhw+05,lwy+16}. 

Table~\ref{tb:position} lists positions from the Nanshan timing analysis for 48 pulsars and also well-defined previously published positions with earlier epochs for these same pulsars. Most of the previously published positions are from papers referenced in the ATNF Pulsar Catalogue V1.54 \citep{mhth05} as listed in the final column of the table. Reference epochs and positions are marked by superscripts "C" and "T", for the Catalogue positions and the timing positions respectively. 
A comparison of the timing positions with the latest published positions from the Catalogue V1.61 shows that our observations have improved the precision of the positions for 24 pulsars, marked by superscript "I" on the pulsar J2000 name.

Table~\ref{tb:proper motion} gives the proper motions determined from Nanshan timing (superscript T), previously published proper motions from V1.61 of the Pulsar Catalogue (C), and proper motions computed by comparison of Nanshan timing positions with earlier published positions as listed in  Table~\ref{tb:position} (CT). We have determined the proper motions for nine pulsars with no previously published proper motions, marked by superscript "N" on the pulsar J2000 name, and improved the precision of the proper motion for a further 13 pulsars, marked by superscript "I". 

PSR J1722$-$3632 had a large glitch (see Sec.~\ref{sec:glitch}). Because of the short span of post-glitch data, we used only the pre-glitch data to fit its position and proper motion. Because PSRs J0629+2415, J1733$-$2228, J1750$-$2438, J1759$-$1940, J1759$-$2922, J1808$-$2057 lie near the ecliptic plane, the positions and proper motions in declination cannot be determined precisely from timing and we only give the proper motions in right ascension. For the other pulsars, our results are consistent with previous measurements.

For many pulsars, the best proper motions are obtained from the position comparison method. Most of of the literature positions are from the \citet{hlk+04} analysis of Jodrell Bank timing data and typically have a position epoch of mid-1994, whereas the Nanshan timing positions have an epoch of mid-2008, giving an effective baseline of about 14 years. For some pulsars, the effective baseline is much larger, up to 30 years. In contrast, the Nanshan timing data span of 12 years gives an effective baseline of six years for proper motion determinations. Consequently, the Nanshan timing proper motions are often less precise than those from the position comparison method. Using the position comparison method, proper motions have been improved for a further eight pulsars, marked by "P" in Table~\ref{tb:proper motion}.

\begin{table*}
\scriptsize
\caption{Positions for 48 pulsars in J2000 equatorial coordinates. Superscripts "C" and "T" indicate values from the literature and the analysis of the Nanshan timing data, respectively. Superscript "I" on the J2000 name indicates an improved measurement. References for the  positions are listed in the last column. Uncertainties in parentheses are in the last quoted digit and are $1\sigma$. }
\centering
\begin{tabular}{lllllllll}
\hline
PSR Name       &PSR Name  &Epoch$^{\rm C}$  &RA$^{\rm C}$    &Dec.$^{\rm C}$         &Epoch$^{\rm T}$ &RA$^{\rm T}$        &Dec.$^{\rm T}$         & Ref.   \\
   (J2000) & (B1950) & (MJD)    &{(h m s)}       &{(d m s)}          & (MJD)    &{(h m s)}          &{(d m s)}          &        \\
\hline
J0055+5117 &   B0052+51 &  49664 &    00:55:45.378(10)  &  +51:17:24.9(1)    &    54611  &   00:55:45.394(3)    &    +51:17:24.66(4)    & 6\\
J0134$-$2937 & $-$      &  50846 &    01:34:18.6690(15) &  $-$29:37:16.91(3) &    54607  &   01:34:18.6826(4) & $-$29:37:17.040(7)    & 6 \\
J0157+6212 &   B0154+61 &  49709 &    01:57:49.93(2)    &  +62:12:25.9(2)    &    54594  &   01:57:49.95(2)     &    +62:12:26.4(2)     & 6\\
J0335+4555 &   B0331+45 &  46109 &    03:35:16.643(2)   &  +45:55:52.99(5)   &    54596  &   03:35:16.645(1)    &    +45:55:53.44(3)    & 1\\
J0520$-$2553 &  $-$     &  51216 &    05:20:36.185(3)   &  $-$25:53:12.28(10)  &    53570  &   05:20:36.194(5)  & $-$25:53:12.16(10)   & 6\\
\\
J0629+2415 &   B0626+24 &  43984 &    06:29:05.719(4)   &  +24:15:41.65(6)      &    54595  &   06:29:05.731(3)    &    $-$                & 1\\
J0653+8051$\rm ^{I}$ &   B0643+80 &  46110 &    06:53:14.86(3)    &  +80:51:59.9(1)    &    54593  &   06:53:15.25(1)   & +80:52:00.03(2)    & 1\\
J0754+3231 &   B0751+32 &  48725 &    07:54:40.688(5)   &  +32:31:56.2(2)    &    54584  &   07:54:40.663(7)    &    +32:31:55.9(3)     & 6\\
J0908$-$1739 & B0906$-$17 &  48737 &    09:08:38.182(1)   &  $-$17:39:37.67(3)   &    54601  &   09:08:38.209(1)  & $-$17:39:39.09(3)  & 6\\
J0944$-$1354 & B0942$-$13 &  49337 &    09:44:28.956(1)   &  $-$13:54:41.63(2)   &    54319  &   09:44:28.962(2) &  $-$13:54:41.84(4)   & 6\\
\\
J1047$-$3032 &    $-$     &  51019 &    10:47:00.815(6)    &  $-$30:32:18.0(1)    &    53571  &   10:47:00.84(3)     & $-$30:32:17.8(5)  & 6\\
J1311$-$1228$\rm ^{I}$ & B1309$-$12 &  49667 &    13:11:52.66(2)    &  $-$12:28:00.7(5)    &    54371  &   13:11:52.656(9)    & $-$12:28:01.1(3)  & 1\\
J1527$-$3931  & B1524$-$39 &  43558 &    15:27:58.93(8)    &  $-$39:31:35(3)      &    54217  &   15:27:58.85(5)     & $-$39:31:35(1)    & 1 \\
J1543$-$0620  & B1540$-$06 &  43889 &    15:43:30.170(8)   &  $-$06:20:45.29(9)   &    54601  &   15:43:30.140(1)    & $-$06:20:45.42(6) & 1\\
J1603$-$2531 &    $-$     &  50719 &    16:03:04.893(3)   &  $-$25:31:47.36(7)    &    54602  &   16:03:04.866(3)  & $-$25:31:47.1(2)  & 6\\
\\
J1614+0737$\rm ^{I}$ &   B1612+07 &  49897 &    16:14:40.906(9)   &  +07:37:31.0(2)    &    54192  &   16:14:40.902(4)  & +07:37:31.7(1)      & 6\\
J1708$-$3426$\rm ^{I}$ &  $-$       &  50856 &    17:08:57.79(4)    &  $-$34:26:44(2)      &    54565  &   17:08:57.784(2)  & $-$34:26:44.8(1)    & 6\\
J1722$-$3632$\rm ^{I}$&    B1718$-$36 &  48378 &    17:22:09.817(8)  & $-$36:32:53.3(5)    &  53882  &   17:22:09.793(2)   &  $-$36:32:55.0(1)    & 1\\
J1733$-$2228$\rm ^{I}$ &   B1730$-$22 &  42004 &    17:33:26(2)    &  $-$22:28:45(45)   &  54132  &   17:33:26.429(9)   &  $-$                 & 1\\
J1735$-$0724 &   B1732$-$07 &  49887 &    17:35:04.972(1)   &  $-$07:24:52.49(7) &  54602  &   17:35:04.974(1)  & $-$07:24:52.14(8)    & 6\\
\\
J1739$-$2903$\rm ^{I}$ &   B1736$-$29 & 49448 &  17:39:34.278(4)   &  $-$29:03:03.5(5)    &    54598  &   17:39:34.289(1)  &    $-$29:03:02.3(2)  & 6\\
J1743$-$3150 &   B1740$-$31 & 48378 &  17:43:36.71(2)    &  $-$31:50:20(2)      &    54599  &   17:43:37.0(1)    &    $-$31:50:22.3(10) & 1\\
J1750$-$2438 &  $-$       &  51491 &   17:50:59.787(9)   &  $-$24:38:58(7)      &    54540  &   17:50:59.780(9)  &    $-$               & 6\\
J1759$-$1940 &    $-$     &  51491 &   17:59:57.04(1)    &  $-$19:40:29(5)      &    53575  &   17:59:57.05(2)   &    $-$               & 6\\
J1759$-$2922$\rm ^{I}$ &$-$ &  50856 &   17:59:48.245(11)    &  $-$29:22:07(3)      &    54589  &   17:59:48.253(7)  &    $-$               & 6\\
\\
J1808$-$2057$\rm ^{I}$ & B1805$-$20 &  49612 &    18:08:06.40(1)    &  $-$20:58:08(3)      &    54599  &   18:08:06.394(6) &    $-$               & 6\\
J1817$-$3618$\rm ^{I}$ & B1813$-$36 &  43558 &    18:17:05.78(2)    &  $-$36:18:04.0(6)    &    54520  &   18:17:05.830(7) &    $-$36:18:04.5(3)  & 6\\
J1820$-$1346$\rm ^{I}$ & B1817$-$13 &  49609 &    18:20:19.76(1)    &  $-$13:46:15(1)      &    54607  &   18:20:19.743(9) &    $-$13:46:15.0(7)  & 6\\
J1823$-$3106$\rm ^{I}$ & B1820$-$31 &  50093 &    18:23:46.788(1)   &  $-$31:06:49.8(1)    &    54607  &   18:23:46.7998(8) & $-$31:06:49.56(7)  & 6\\
J1836$-$0436$\rm ^{I}$ & B1834$-$04 &  49532 &    18:36:51.790(2)   &  $-$04:36:37.7(1)    &    54515  &   18:36:51.788(1)  & $-$04:36:37.69(6) & 6\\
\\
J1839$-$0643$\rm ^{I}$ & $-$     &  51368 &    18:39:09.80(1)    &  $-$06:43:45(2)    &    54603  &   18:39:09.788(7)    &    $-$06:43:44.5(4)   & 3\\
J1844+1454 &   B1842+14 &  49362 &    18:44:54.895(1)   &  +14:54:14.12(2)   &    54602  &   18:44:54.897(2)  & +14:54:14.59(3)     & 6\\
J1857+0526$\rm ^{I}$ &   $-$      &  51800 &    18:57:15.856(8)   &  +05:26:28.7(5)    &    54619  &   18:57:15.851(2)    &    +05:26:28.64(7)    & 5\\
J1901+0716$\rm ^{I}$ &   B1859+07 &  49863 &    19:01:38.94(2)    &  +07:16:34.8(4)    &    54537  &   19:01:38.956(5)  & +07:16:34.4(1)     & 6\\
J1902+0556$\rm ^{I}$ &   B1900+05 &  49722 &    19:02:42.620(3)   &  +05:56:25.9(1)    &    54288  &   19:02:42.616(3)  & +05:56:25.92(6)      & 6\\
\\
J1903+0135$\rm ^{I}$ &   B1900+01 &  42345 &    19:03:29.976(2)   &  +01:35:38.20(7)  &   54587  &  19:03:29.983(1)  & +01:35:38.39(3) & 1\\
J1905+0709 &   B1903+07 &  49466 &    19:05:53.62(2)    &  +07:09:19.4(6)    &    54219  &   19:05:53.62(2)   & +07:09:19.2(6)   & 6\\
J1916+0951$\rm ^{I}$ &   B1914+09 &  42824 &    19:16:32.342(4)   &  +09:51:25.29(7)   &    54587  &   19:16:32.3391(9) & +09:51:25.93(2)    & 1\\
J1918+1444 &   B1916+14 &  49690 &    19:18:23.638(6)   &  +14:45:06.0(1)    &    54644  &   19:18:23.679(10)   &    +14:45:05.8(2)     & 6\\
J1926+0431$\rm ^{I}$ &   B1923+04 &  48716 &    19:26:24.470(9)   &  +04:31:31.6(2)    &    54606  &   19:26:24.471(3)  & +04:31:31.37(8)     & 6\\
\\
J1937+2544 &   B1935+25 &  49703 &    19:37:01.266(1)   &  +25:44:13.68(2)   &    54248  &   19:37:01.2564(8) & +25:44:13.52(1)    & 6\\
J1943$-$1237$\rm ^{I}$ & B1940$-$12 &  48717 &    19:43:25.481(8)   &  $-$12:37:42.4(5)&    54561  &   19:43:25.480(4)  & $-$12:37:43.1(3)   & 6\\
J2004+3137$\rm ^{I}$ &   B2002+31 &  42303 &    20:04:52.295(4)   &  +31:37:10.19(9)   &    54539  &   20:04:52.279(2)  & +31:37:09.99(3)    & 1\\
J2029+3744$\rm ^{I}$ &   B2027+37 &  49725 &    20:29:23.872(10)  &  +37:44:08.2(1)    &    54150  &   20:29:23.882(6)  & +37:44:08.05(8)     & 6\\
J2055+2209$\rm ^{I}$ &  B2053+21 &  49726 &    20:55:39.151(5) &  +22:09:27.2(1)    &    54546  &  20:55:39.161(2)  & +22:09:27.09(5)    & 6\\
\\
J2113+2754 &   B2110+27 &  48741 &    21:13:04.390(2)   &  +27:54:02.29(3)   &    54595  &   21:13:04.360(3)    &    +27:54:01.39(5)    & 6\\
J2149+6329 &   B2148+63 &  43890 &    21:49:58.59(3)    &  +63:29:43.5(2)    &    54595  &   21:49:58.701(3)    &    +63:29:44.23(2)    & 1\\
J2305+3100 &   B2303+30 &  48714 &    23:05:58.333(4)    &  +31:00:01.80(8)    &    54602  &   23:05:58.322(3)    &    +31:00:01.36(6)    & 1\\
\hline
\end{tabular}
\begin{flushleft}
\scriptsize
The references for Table~\ref{tb:position} and Table~\ref{tb:proper motion} are as follows:
(1)\citet{tml93},
(2)\citet{hla93} ,
(3)\citet{mhl+02} ,
(4)\citet{bfg+03} ,
(5)\citet {kbm+03} ,
(6)\citet{hlk+04},
(7)\citet{lwy+16},
(8)\citet{dgb+19} ,
(9)\citet{jbvk+19}
\end{flushleft}

\end{table*}
\label{tb:position}
\begin{table*}
\caption{Proper motions of 48 pulsars. Superscripts "T", "C" and "CT" indicate values of $\mu_\alpha$ and $\mu_\delta$ from the timing analysis, previously published proper motions and proper motions from comparison of literature and timing positions, respectively. References for the best available previously published proper motions (from the ATNF Catalogue V1.61) are given in the last column. Superscripts "N" and "I" on the J2000 name indicate new and improved proper motions from the Nanshan timing, respectively, and superscript "P" indicates improved proper motions from the position comparison method. Uncertainties in parentheses are in the last quoted digit and are $1\sigma$. }
\centering
\scriptsize
\begin{tabular}{llllllllllc}
\hline
PSR Name &PSR Name    & $\mu_\alpha\;^{\rm T}$ &$\mu_\alpha\;^{\rm C}$ &$\mu_\alpha\;^{\rm CT}$ & $\mu_\delta\;^{\rm T}$ &$\mu_\delta \;^{\rm C}$&$\mu_\delta \;^{\rm CT}$  &Ref.  \\
  (J2000)             &  (B1950)  &{(mas  yr$^{-1}$)}   &{(mas yr$^{-1}$)}     &{(mas  yr$^{-1}$)}    &{(mas yr$^{-1}$)}  &{(mas  yr$^{-1}$)}    &{(mas yr$^{-1}$)}  &  \\
\hline
J0055+5117            &  B0052+51     &   8(10)       &10.49(6)     & 11(7)      &  $-$1(12)       & $-$17.35(2)    & $-$23(8)    & 8      \\  
J0134$-$2937          &  $-$          &   13(1)       &17(1)        & 17(1)      &  $-$12(2)       & $-$9(2)        & $-$13(2)    & 9      \\  
J0157+6212            &   B0154+61    &   $-$11(23)   &1.5(1)       & 9(12)      &  30(30)         & 44.81(4)       & 32(14)      & 8      \\  
J0335+4555            &   B0331+45    &   $-$6(4)     &$-$3.64(7)   & 4(1)       &  $-$6(9)        & $-$0.1(1)      & 19(2)       & 8      \\     
J0520$-$2553          &     $-$       &   53(48)      &$-$1(19)     & 20(12)     &  16(72)         & $-$7(41)       & 19(21)      & 6      \\ 
\\       		    	     	                                                                                             
J0629+2415            &   B0626+24    &   11(13)      &3.6(2)       & 6(2)       &  $-$            & $-$4.6(2)      & $-$         & 8      \\
J0653+8051            &   B0643+80    &   24(9)       &19(3)        & $-$69(4)   &  $-$6(8)        & $-$1(3)        & $-$6(4)     & 2      \\
J0754+3231            &   B0751+32    &   $-$33(28)   &$-$4(5)      & $-$20(7)   &  $-$140(102)    & 7(3)           & $-$17(23)   & 2      \\
J0908$-$1739$\rm ^{I}$ & B0906$-$17    &   18(5)       &27(11)       &24(2)       &  $-$111(8)      & $-$40(11)      & $-$88(3)    & 2      \\ 
J0944$-$1354$\rm ^{I}$ &B0942$-$13     &   11(9)       &$-$1(32)     &6(2)        &  $-$31(13)      & $-$22(14)      & $-$15(3)    & 2      \\ 
\\       		    	     	           	                                                                                      
J1047$-$3032$\rm ^{P}$&     $-$       &   $-$60(243)  &48(66)       & 55(54)     &  $-$373(295)    & $-$44(75)      & 34(69)      & 6      \\ 
J1311$-$1228$\rm ^{I}$& B1309$-$12     &   62(40)      &5(51)        &$-$2(8)     &  129(99)        & 17(117)        & $-$20(26)   & 6      \\
J1527$-$3931$\rm ^{N}$& B1524$-$39     &   $-$206(216) &$-$          &$-$33(37)   &  $-$348(515)    & $-$            & $-$12(111)  & $-$    \\ 
J1543$-$0620          &   B1540$-$06  &   $-$19(6)    &$-$16.77(6)  & $-$15(4)   &  $-$18(19)      & $-$0.3(1)      & $-$4(4)     & 8      \\
J1603$-$2531$\rm ^{I}$&    $-$         &   $-$11(12)   &11(13)       &$-$34(4)    &  145(58)        & 123(60)        & 22(18)      & 6      \\
\\       		    	     	           	                                                                                       
J1614+0737$\rm ^{I}$  &   B1612+07     &   $-$20(25)   &12(32)       &$-$5(12)    &  59(53)         & 37(83)         & 58(23)       & 6      \\
J1708$-$3426$\rm ^{N}$&   $-$          &   $-$31(9)    &$-$          &$-$6(25)    &  123(45)        & $-$            & $-$67(198)   & $-$    \\
J1722$-$3632$\rm ^{N}$& B1718$-$36     &   12(14)      &$-$          &$-$19(7)    &  $-$34(62)      & $-$            & $-$110(34)   & $-$    \\
J1733$-$2228         &   B1730$-$22   &   15(27)      &$-$43(84)    & 71(9)      &  $-$            & $-$700(1800)   & $-$          & 6      \\
J1735$-$0724         &   B1732$-$07   &   $-$3(6)     &$-$2.4(17)   & 2(2)       &  9(23)          & $-$28(3)       & 27(8)        & 8      \\       		    
\\                                                                                                                                 
J1739$-$2903$\rm ^{N}$& B1736$-$29     &   9(7)        &$-$          &10(4)       &  115(63)        & $-$            & 84(41)       & $-$    \\ 
J1743$-$3150$\rm ^{I}$& B1740$-$31     &   $-$60(46)   &$-$36(61)    &$-$11(17)   &  $-$924(317)    & 100(600)       & $-$136(131)  & 6      \\
J1750$-$2438$\rm ^{N}$&  $-$           &   $-$18(48)   &$-$          &$-$11(21)   &  $-$            & $-$            & $-$          & $-$    \\
J1759$-$1940$\rm ^{N}$&  $-$           &   247(208)    &$-$          &24(70)      &  $-$            & $-$            & $-$          & $-$    \\
J1759$-$2922         &     $-$        &   $-$8(26)    &$-$3(59)     & 10(12)     &  $-$            & $-$0(1100)     & $-$          & 6      \\
\\                                                                                                                                 
J1808$-$2057          &   B1805$-$20  &   $-$25(25)   &2(36)        & $-$2(12)   &  $-$            & $-$1200(700)   & $-$          & 6      \\ 
J1817$-$3618          &  B1813$-$36   &   $-$26(27)   & 19(5)       & 21(6)      &  43(109)        & $-$16(17)      & $-$17(23)    & 9    \\
J1820$-$1346$\rm ^{I}$&B1817$-$13      &   4(37)       &$-$15(38)    &$-$17(16)   &  $-$36(215)     & 100(300)       & 21(95)       & 6      \\ 
J1823$-$3106$\rm ^{P}$& B1820$-$31    &   7(4)        &18(4)        & 12(1)      &  $-$16(26)      & 27(18)         & 15(11)       & 9      \\ 
J1836$-$0436$\rm ^{P}$&   B1834$-$04  &   $-$3(7)     &0(6)         & $-$3(2)    &  1(22)          & 16(19)         & $-$1(9)      & 6      \\
\\                                                                                                                                 
J1839$-$0643$\rm ^{N}$ & $-$           &   4(32)       &$-$          &$-$21(21)   &  $-$170(111)    & $-$            & 61(121)     & $-$    \\
J1844+1454$\rm ^{P}$ & B1842+14    &   19(8)       &$-$9(10)     & 3(2)       &  60(12)         & $-$45(6)       & 33(3)       & 4      \\
J1857+0526$\rm ^{N}$   &  $-$          &   19(11)      &$-$          &$-$9(16)    &  $-$51(20)      & $-$            & $-$8(65)    & $-$    \\
J1901+0716$\rm ^{I}$   &   B1859+07    &   26(26)      &$-$45(43)    &23(16)      &  $-$121(48)     & $-$9(89)       & $-$28(33)   & 6      \\ 
J1902+0556$\rm ^{P}$  &   B1900+05    &   $-$19(15)   &$-$7(13)     & $-$4(5)    &  31(24)         & $-$4(29)       & $-$1(10)     & 6      \\
\\                                                                                                                                 
J1903+0135$\rm ^{I}$   &   B1900+01   &   11(6)       &3(7)         & 3(1)        &  $-$7(12)       & $-$13(14)      & 6(2)        & 7      \\
J1905+0709$\rm ^{P}$   &   B1903+07   &   $-$47(107)  &$-$58(70)    &  $-$1(34)   &  $-$236(204)    & 57(148)        & $-$12(63)   & 6      \\ 
J1916+0951$\rm ^{I}$  &  B1914+09     &   $-$14(5)    &$-$9(7)      & $-$1(2)     &  $-$3(9)        & $-$1(13)       & $-$11(2)    & 6      \\ 
J1918+1444$\rm ^{N}$  &  B1916+14     &   78(49)      &$-$          & 44(13)      &  $-$262(77)     & $-$            & $-$12(20)   & $-$    \\
J1926+0431$\rm ^{I}$  &  B1923+04     &   $-$2(18)    &$-$4(20)     & 1(9)        &  $-$44(33)      & 22(48)         & $-$14(16)   & 6      \\
\\                                                                                                                                 
J1937+2544            &  B1935+25     &   $-$21(4)    &$-$10.05(4)  & $-$10(2)   &  $-$10(5)       & $-$13.06(4)    & $-$12(2)    & 9      \\
J1943$-$1237$\rm ^{P}$ &   B1940$-$12  &   $-$30(25)   &$-$20(23)    & $-$1(8)    &  $-$88(97)      & $-$51(135)     & $-$46(35)   & 6      \\
J2004+3137$\rm ^{P}$ &   B2002+31    &   $-$10(10)   &0(9)         & $-$6(2)    &  $-$4(11)       & $-$9(13)       & $-$6(3)     & 6      \\
J2029+3744$\rm ^{I}$  &  B2027+37     &   $-$8(28)    &$-$25(31)    & 10(11)      &  74(32)         & 15(34)         & $-$10(13)   & 6      \\
J2055+2209$\rm ^{I}$  &  B2053+21     &   $-$3(10)    &$-$9(15)     & 10(5)       &  4(14)          & $-$5(22)       & $-$10(9)    & 6      \\
\\                                                                                                                                 
J2113+2754            &   B2110+27    &   $-$35(13)   &$-$27.98(5)  & $-$24(3)   &  $-$74(15)      & $-$54.43(9)    & $-$56(4)    & 8      \\
J2149+6329            &   B2148+63    &   12(6)       &15.8(1)      & 26(7)      &  21(6)          & 11.3(3)        & 25(7)       & 8      \\
J2305+3100            &   B2303+30    &   4(13)       &$-$3.74(8)   & $-$4(2)    &  $-$40(18)      & $-$15.6(2)     & $-$13(3)    & 8      \\
\hline
\end{tabular}
\end{table*}
\label{tb:proper motion}

\subsection{Velocities}\label{sec:Velocities}

In general, if we know the proper motion and distance of a pulsar, its transverse velocity is given by $V_{T} = 4.74\;\mu_{\rm tot}D$~km~s$^{-1}$, where $\mu_{\rm tot} = \sqrt{\mu_\alpha^2+\mu_\delta^2}$ is the total proper motion in mas~yr$^{-1}$, and $D$ is the pulsar distance in kpc. Table~3 lists the derived velocities for 48 pulsars. Distances ($D$) obtained from the YMW16 \citep{ymw16} and NE2001 \citep{cl02} Galactic electron-density distribution models and from VLBI observations \citep{dgb+19} 
are given in columns 3 -- 5, respectively. The next four columns give the total proper motion ($\mu_{\rm tot}$) from pulsar timing and the transverse velocities ($V_T$) corresponding to the three distance estimates. Finally, the last four columns give the same results based on the position-comparison proper motions. Uncertainties in pulsar velocities are computed using standard error propagation and assuming a 10\% uncertainty in the estimated distances.
Note that, for the YMW16 distance model, PSRs J0134$-$2937, J1311$-$1228 and J2305+3100 have DMs higher than the maximum Galactic DM in this direction. In these cases, the YMW16 model sets a distance of 25kpc by default. We did not use these distances to calculate the transverse velocities. For six pulsars that are located near the ecliptic plane, we give one-dimensional velocities, $\rm V_{1D}$ in Table ~\ref{tab:1D_velocities}, where $V_{\rm 1D} = 4.74\;\mu_\alpha\;D$, where $\mu_\alpha$ is the proper motion in the right ascension direction as in Table ~\ref{tb:proper motion}.

\begin{longrotatetable}
\begin{deluxetable*}{lllllllllllll}
\centering
\tablecaption{Velocities of 48 pulsars.  The subscripts "Y", "CL" and "D" indicate distances obtained from \citet{ymw16}, \citet{cl02} and \citet{dgb+19}, respectively. The superscripts "T" and "CT" indicate proper motions obtained from the Nanshan timing and comparison of the Catalogue and timing positions, respectively.}
\label{tb:2D_velocities}
\tablewidth{600pt}
\tabletypesize{\footnotesize}
\tablehead{
PSR Name  &PSR Name    & Dist.$_{\rm Y}$& Dist.$_{\rm CL}$  &Dist.$_{\rm D}$ & $\mu_{\rm tot}^{\rm T}$ & V$_{\rm Y}^{\rm T}$ & V$_{\rm CL}^{\rm T}$ & V$_{\rm D}^{\rm T}$ & $\mu_{\rm tot}^{\rm CT}$ & V$_{\rm Y}^{\rm CT}$ & V$_{\rm CL}^{\rm CT}$ & V$_{\rm D}^{\rm CT}$    \\
  (J2000)   & (B1950)   & (kpc) & (kpc) & (kpc)& {(mas yr$^{-1}$)}   & {(km s$^{-1}$)} & {(km s$^{-1}$)} & {(km s$^{-1}$)}  & {(mas yr$^{-1}$)}   & {(km s$^{-1}$)} & {(km s$^{-1}$)} & {(km s$^{-1}$)}
}
\startdata
J0055+5117     &  B0052+51     &    1.94   &  1.9   & 2.87$^{+0.54}_{-0.39}$&  8(10)      &  78(94)         &  74(92)        &  112(139)    &  25(8)       & 230(86)       & 230(84)        &  340(127)     \\
J0134$-$2937   &  $-$          &    25.0   &  0.56  & $-$                   &  18(2)      &  $-$            &  48(11)        &  $-$	    &  21(1)       & $-$           & 56(12)         &  $-$          \\
J0157+6212     &  B0154+61     &    1.39   &  1.71  & 1.8$^{+0.08}_{-0.12}$ &  31(29)     &  200(196)       &  250(240)      &  260(250)    &  34(14)      & 220(103)      & 270(126)       &  290(133)     \\
J0335+4555     &  B0331+45     &    1.54   &  1.64  & 2.44$^{+0.18}_{-0.12}$&  9(7)       &  62(50)         &  66(53)        &  99(79)      &  20(2)       & 145(34)       & 154(36)        &  230(54)      \\
J0520$-$2553   &  $-$          &    2.38   &  1.74  & $-$                   &  56(50)     &  630(580)       &  460(430)      &  $-$	    &  27(17)      & 310(202)      & 220(148)       &  $-$          \\
\\                                                                                                                                                                                                             
J0629+2415     &  B0626+24     &    1.67   &  2.24  & 3.0$^{+0.57}_{-0.29}$ &  $-$        &  $-$            &  $-$           &  $-$	    &  $-$         & $-$           & $-$            &  $-$          \\
J0653+8051     &  B0643+80     &    2.35   &  1.55  & $-$                   &  24(9)      &  270(112)       &  179(74)       &  $-$	    &  69(4)       & 770(159)      & 510(105)       &  $-$          \\
J0754+3231     &  B0751+32     &    1.46   &  1.52  & $-$                   &  144(99)    &  1000(720)      &  1000(750)     &  $-$	    &  26(16)      & 181(117)      & 189(122)       &  $-$          \\
J0908$-$1739   &  B0906$-$17   &    0.8    &  0.91  & $-$                   &  112(8)     &  430(90)        &  490(102)      &  $-$	    &  92(2)       & 350(70)       & 400(80)        &  $-$          \\
J0944$-$1354   &  B0942$-$13   &    0.69   &  0.67  & $-$                   &  33(13)     &  108(47)        &  105(46)       &  $-$	    &  16(3)       & 54(15)        & 52(15)         &  $-$          \\
\\                                                                                                                                                                                                             
J1047$-$3032   &  $-$          &    2.15   &  2.37  & $-$                   &  380(300)   &  3800(3100)     &  4200(3400)    &  $-$	    &  65(58)      & 660(608)      & 730(670)       &  $-$          \\
J1311$-$1228   &  B1309$-$12   &    25.0   &  3.14  & $-$                   &  143(91)    &  $-$            &  2100(1400)    &  $-$	    &  20(26)      & $-$           & 300(387)       &  $-$          \\
J1527$-$3931   &  B1524$-$39   &    1.71   &  1.51  & $-$                   &  410(460)   &  3300(3800)     &  2900(3300)    &  $-$	    &  35(51)      & 280(414)      & 250(365)       &  $-$          \\
J1543$-$0620   &  B1540$-$06   &    1.12   &  0.72  & 3.11$^{+0.51}_{-0.25}$&  27(14)     &  142(78)        &  91(50)        &  400(200)    &  16(4)       & 84(27)        & 54(18)         &  230(76)      \\
J1603$-$2531   &  $-$          &    3.41   &  1.87  & $-$                   &  145(58)    &  2300(1100)     &  1300(580)     &  $-$	    &  41(10)      & 660(210)      & 360(115)       &  $-$          \\
\\                                                                                                                                                                                                             
J1614+0737     &  B1612+07     &    1.63   &  0.96  & $-$                   &  63(51)     &  480(410)       &  280(240)      &  $-$	    &  58(23)      & 450(201)      & 270(119)       &  $-$          \\
J1708$-$3426   &  $-$          &    4.73   &  3.56  & $-$                   &  85(70)     &  1900(1600)     &  1400(1200)    &  $-$	    &  67(196)     & 1500(4400)    & 1100(3300)     &  $-$          \\
J1722$-$3632   &  B1718$-$36   &    3.99   &  4.36  & $-$                   &  39(66)     &  750(1300)      &  810(1400)     &  $-$	    &  114(34)     & 2200(780)     & 2300(850)      &  $-$          \\
J1733$-$2228   &  B1730$-$22   &    1.1    &  1.17  & $-$                   &  $-$        &  $-$            &  $-$           &  $-$	    &  $-$         & $-$           & $-$            &  $-$          \\
J1735$-$0724   &  B1732$-$07   &    0.21   &  2.26  & 6.68$^{+2.03}_{-1.43}$&  9(22)      &  9(22)          &  100(230)      &  280(690)    &  27(8)       & 27(10)        & 290(104)       &  850(300)     \\
\\                                                                                                                                                                                                             
J1739$-$2903   &  B1736$-$29   &    2.91   &  2.47  & $-$                   &  116(62)    &  1600(920)      &  1300(800)     &  $-$	    &  84(40)      & 1200(600)     & 990(510)       &  $-$          \\
J1743$-$3150   &  B1740$-$31   &    3.33   &  3.31  & $-$                   &  930(320)   &  14600(5800)    &  14500(5800)   &  $-$	    &  137(131)    & 2200(2100)    & 2100(2100)     &  $-$          \\
J1750$-$2438   &  $-$          &    12.9   &  6.58  & $-$                   &  $-$        &  $-$            &  $-$           &  $-$	    &  $-$         & $-$           & $-$            &  $-$          \\
J1759$-$1940   &  $-$          &    7.98   &  4.92  & $-$                   &  $-$        &  $-$            &  $-$           &  $-$	    &  $-$         & $-$           & $-$            &  $-$          \\
J1759$-$2922   &  $-$          &    2.42   &  1.96  & $-$                   &  $-$        &  $-$            &  $-$           &  $-$	    &  $-$         & $-$           & $-$            &  $-$          \\
\\                                                                                                                                                                                                             
J1808$-$2057   &  B1805$-$20   &    4.58   &  7.61  & $-$                   &  $-$        &  $-$            &  $-$           &  $-$	    &  $-$         & $-$           & $-$            &  $-$          \\
J1817$-$3618   &  B1813$-$36   &    4.4    &  2.44  & $-$                   &  50(95)     &  1100(1200)     &  600(1100)     &  $-$	    &  27(15)      & 570(336)      & 320(186)       &  $-$          \\
J1820$-$1346   &  B1817$-$13   &    5.89   &  8.83  & $-$                   &  36(220)    &  1000(6000)     &  1500(8900)    &  $-$	    &  27(75)      & 760(2100)     & 1100(3100)     &  $-$          \\
J1823$-$3106   &  B1820$-$31   &    1.59   &  1.29  & $-$                   &  17(23)     &  130(178)       &  106(148)      &  $-$	    &  19(8)       & 145(69)       & 118(56)        &  $-$          \\
J1836$-$0436   &  B1834$-$04   &    4.36   &  4.94  & $-$                   &  3(9)       &  60(177)        &  68(200)       &  $-$	    &  3(3)        & 58(71)        & 66(80)         &  $-$          \\
\\                                                                                                                                                                                                             
J1839$-$0643   &  $-$          &    4.82   &  6.65  & $-$                   &  170(111)   &  3800(2600)     &  5300(3700)    &  $-$	    &  65(114)     & 1500(2600)    & 2000(3600)     &  $-$          \\
J1844+1454     &  B1842+14     &    1.68   &  2.19  & $-$                   &  63(12)     &  500(137)       &  650(178)      &  $-$	    &  33(3)       & 260(58)       & 340(76)        &  $-$          \\
J1857+0526     &  $-$          &    12.27  &  8.8   & $-$                   &  54(19)     &  3100(1300)     &  2200(1000)    &  $-$	    &  12(46)      & 720(2700)     & 510(1900)      &  $-$          \\
J1901+0716     &  B1859+07     &    3.4    &  5.47  & $-$                   &  124(48)    &  2000(870)      &  3200(1400)    &  $-$	    &  36(27)      & 580(450)      & 940(730)       &  $-$          \\
J1902+0556     &  B1900+05     &    3.6    &  4.66  & $-$                   &  37(22)     &  630(390)       &  810(500)      &  $-$	    &  4(5)        & 73(92)        & 95(119)        &  $-$          \\
\\                                                                                                                                                                                                             
J1903+0135     &  B1900+01     &    3.3    &  3.33  & $-$                   &  13(8)      &  210(131)       &  200(132)      &  $-$	    &  7(2)        & 102(38)       & 103(39)        &  $-$          \\
J1905+0709     &  B1903+07     &    4.98   &  5.67  & $-$                   &  240(200)   &  5700(4900)     &  6500(5600)    &  $-$	    &  12(63)      & 280(1500)     & 320(1700)      &  $-$          \\
J1916+0951     &  B1914+09     &    1.9    &  2.94  & $-$                   &  14(5)      &  125(54)        &  193(84)       &  $-$	    &  11(2)       & 102(29)       & 158(45)        &  $-$          \\
J1918+1444     &  B1916+14     &    1.3    &  1.94  & $-$                   &  270(75)    &  1700(600)      &  2500(900)     &  $-$	    &  1600(94)    & 9800(2000)    & 14500(3000)    &  $-$          \\
J1926+0431     &  B1923+04     &    4.99   &  3.8   & $-$                   &  44(33)     &  1000(800)      &  780(600)      &  $-$	    &  14(16)      & 340(390)      & 260(300)       &  $-$          \\
\\                                                                                                                                                                                                             
J1937+2544     &  B1935+25     &    2.87   &  3.25  & 3.15$^{+0.32}_{-0.28}$&  24(4)      &  320(88)        &  370(99)       &  355(96)     &  16(2)       & 220(51)       & 250(57)        &  240(56)      \\
J1943$-$1237   &  B1940$-$12   &    1.2    &  1.21  & $-$                   &  93(92)     &  530(530)       &  530(540)      &  $-$	    &  46(35)      & 260(200)      & 270(200)       &  $-$          \\
J2004+3137     &  B2002+31     &    8.0    &  7.5   & $-$                   &  11(10)     &  430(380)       &  340(360)      &  $-$	    &  8(2)        & 310(109)      & 300(102)       &  $-$          \\
J2029+3744     &  B2027+37     &    5.77   &  6.33  & $-$                   &  75(32)     &  2000(960)      &  2200(1100)    &  $-$	    &  14(12)      & 380(340)      & 420(370)       &  $-$          \\
J2055+2209     &  B2053+21     &    3.0    &  2.43  & $-$                   &  5(13)      &  75(180)        &  61(146)       &  $-$	    &  14(7)       & 200(110)      & 165(89)        &  $-$          \\
\\                                                                                                                                                                                                             
J2113+2754     &  B2110+27     &    1.87   &  2.03  & 1.42$^{+0.04}_{-0.04}$&  82(15)     &  730(197)       &  790(200)      &  550(150)    &  61(4)       & 540(113)      & 590(123)       &  410(86)      \\
J2149+6329     &  B2148+63     &    3.88   &  5.62  & 2.81$^{+0.58}_{-0.47}$&  24(6)      &  440(146)       &  640(200)      &  320(107)    &  36(7)       & 660(183)      & 960(270)       &  480(133)     \\
J2305+3100     &  B2303+30     &    25.0   &  3.67  & 4.47$^{+0.65}_{-0.58}$&  40(18)     &  $-$            &  700(350)      &  850(430)    &  14(3)       & $-$           & 240(69)        &  290(84)      \\		
\\
\enddata
\end{deluxetable*}
\end{longrotatetable}

\begin{table*}
\caption{One-dimensional velocities of pulsars near the ecliptic plane. }
\normalsize
\centering
\label{tab:1D_velocities}
\begin{tabular}{lllllllll}
\hline
\hline
PSR Name      & PSR Name  & ${\mu_\alpha}^{\rm T}$   & Dist.$_{\rm Y}$ & Dist.$_{\rm CL}$ & Dist.$_{\rm D}$ &  V$_{\rm 1D}^{\rm Y}$ & V$_{\rm 1D}^{\rm CL}$ & V$_{\rm 1D}^{\rm D}$   \\
(J2000)& (B1950) & (mas yr$^{-1}$) &   (kpc)   & (kpc)     & (kpc)    &  {(km s$^{-1}$)}     &  {(km s$^{-1}$)}     &  {(km s$^{-1}$)}     \\
\hline
J0629+2415   &   B0626+24   & 11(13)    & 1.67 &  2.24 & 3.0$^{+0.57}_{-0.29}$ & 86(100)    &  115(140)   & 154(190) \\
J1733$-$2228 &   B1730$-$22   & 15(27)    & 1.1  &  1.17 & $-$ & 76(140)    &  81(150)    & $-$   \\
J1750$-$2438 &   $-$        & $-$18(48) & 12.9 &  6.58 & $-$ & 1100(2900) &  580(1500)  & $-$   \\
J1759$-$1940 &   $-$        & 250(210)  & 7.98 &  4.92 & $-$ & 9400(8100) &  5800(5000) & $-$   \\
J1759$-$2922 &   $-$        & $-$8(26)  & 2.42 &  1.96 & $-$ & 91(300)    &  74(240)    & $-$   \\
J1808$-$2057 &   B1805$-$20 & $-$25(25) & 4.58 &  7.61 & $-$ & 530(550)   &  880(920)   & $-$   \\
\hline
\end{tabular}
\end{table*}

\subsection{Rotational parameters of 84 pulsars}\label{sec:Rotational_parameters}
The rotational parameters ($\nu$, $\dot{\nu}$ and $\ddot{\nu}$) of 84 pulsars from full timing analyses, including modelling of the red noise, are presented in Table~\ref{tb:parameters}. The first and second columns are the pulsar names based on the J2000 and B1950 coordinates, respectively. The third to the seventh columns are the reference MJD for $\nu$, $\dot{\nu}$, $\ddot{\nu}$ (where its magnitude is greater than the uncertainty) and braking indices (The uncertainties are from the error propagation formula.), respectively. The next four columns are, in turn, the MJD range of the data, the data span in years, the weighted root-mean-square ($W_{\rm rms}$) of the timing residuals and the number of ToAs. The final three columns give parameters from the red-noise analysis: $f_{c}$, $\alpha$ and $A$. The epoch of the timing model is set to an integral MJD near the center of the data span. Three pulsars suffered glitches during the observed data span. Timing parameters for these pulsars are given in Section~\ref{sec:glitch} below. For these analyses, positions and proper motions from the Nanshan timing were used for the 48 pulsars listed in Tables~\ref{tb:position} and \ref{tb:proper motion}. For the remaining pulsars, positions and proper motions were obtained from the Pulsar Catalogue, V1.61. 

Compared with previously published values, the measured timing parameters have smaller uncertainties for 36 pulsars (including the three glitch pulsars), marked by superscript "I" on the pulsar J2000 name, and others are consistent with previous work. The timing properties of the observed sample of pulsars are further discussed in Section~\ref{sec:discussion}.
\begin{longrotatetable}
\begin{deluxetable*}{lllllllllllllllll}
\centering
\tablecaption{Rotational parameters ($\nu$, $\dot{\nu}$ and $\ddot{\nu}$) of 84 pulsars without glitches. Uncertainties for the rotation parameters are 1$\sigma$ from Tempo2 and for the braking indices are from the error propagation formula in the last quoted digit are given in parentheses.\label{tb:parameters}}
\tablewidth{600pt}
\tabletypesize{\scriptsize}
\tablehead{
PSR Name      & PSR Name       &   Epoch  &  $\nu$             & $\dot{\nu}$         &  $\ddot{\nu}$   & n  &  Data Range   & $\rm T_{s}$  & $\rm W_{rms}$     & N$_{\rm ToA}$ & $f_{\rm c}$  & $\rm \alpha$ & A \\
   (J2000)   &  (B1950)    &          &  (s$^{-1}$)        & ($\rm 10^{-15}s^{-2}$)  &($\rm 10^{-25}s^{-3}$) & &  (MJD)        & (yr)       &($\mu$s) &           & (yr$^{-1})$    &         & ($\rm 10^{-21}yr^{-3}$)
   }
\startdata
J0014+4746$\rm ^{I}$   &  B0011+47    &    54602  &  0.805997070204(2)  &   $-$0.36671(4)   &  ---          & ---  	    	  &  52485-56719  &    11.6   &  4456  &419  &  0.12  &  $-$0.1   &  1.262   \\
J0055+5117      &  B0052+51    &    55042  &  0.4727740626927(6) &   $-$2.131596(5)  &  $-$0.008(1)  & $-$91(14)    	  &  52502-56719  &    11.5   &  947   &391  &  0.1   &  $-$3.5   &  3.102   \\
J0108$-$1431$\rm ^{I}$ &  ---         &    54611  &  1.238290950836(5)  &   $-$0.11867(6)   &  ---          & ---  	    	  &  52502-56719  &    11.5   &  3232  &364  &  0.12  &  $-$1.5   &  20.43  \\
J0134$-$2937    &  ---         &    54607  &  7.3013159483149(9) &   $-$4.177821(9)  &  0.007(3)     & 260(150)     	  &  52494-56719  &    11.6   &  82    &233  &  0.12  &  $-$1     &  0.002   \\
J0151$-$0635    &  B0148$-$06  &    54492  &  0.682750088094(2)  &   $-$0.20764(3)   &  ---          & ---  	    	  &  52471-56513  &    11.1   &  1641  &182  &  0.15  &  $-$1     &  5.309   \\
\\
J0157+6212$\rm ^{I}$   &  B0154+61    &    54594  &  0.425201758139(2)  &   $-$34.14773(2)  &  0.136(7)     & 4.8(2)       	  &  52470-56719  &  11.6     &  579     & 542 &  0.15 &   $-$6   & 4302  \\
J0304+1932      &  B0301+19    &    54594  &  0.720676539013(1)  &   $-$0.67303(2)   &  0.021(4)     & 3230(630)          &  52471-56717  &  11.6     &  1591    &394  &  0.1  &   $-$1   & 1.071     \\
J0335+4555      &  B0331+45    &    54596  &  3.714702689639(2)  &   $-$0.10146(2)   &  ---          & ---                &  52473-56719  &  11.6     &  363     &348  &  0.1  &   $-$1   & 0.004     \\
J0450$-$1248    &  B0447$-$12  &    54596  &  2.283030991909(10) &   $-$0.53537(6)   &  ---          & ---             	  &  52473-56719  &  11.6     &  1910    &386  &  0.12 &   $-$0.1 & 0.518    \\
J0520$-$2553    &  ---         &    53570  & 4.13835031869(2)    &  $-$0.5137(4)     &  $-$0.06(3)   & $-$990000(350000)  &  52486-54654  &  5.9      &  1088    &111  &  0.1  &   $-$1   & 0.444    \\
\\
J0538+2817      &  ---         &    55513  &  6.98520770574(1)   &   $-$179.309(2)   &  $-$0.61(2)   & $-$133(5)  	  &  54604-56423  &    5.8   &  390    &227  & 0.12 &   $-$5     & 1066    \\
J0614+2229$\rm ^{I}$   &  B0611+22    &    54595  &  2.9852067152(1)    &   $-$52.7267(1)   &  53.8(6)      & 56.6(6)  	  &  52473-56718  &    11.6  &  75193  &917  & 0.12 &   $-$6     & 15828900   \\
J0629+2415      &  B0626+24    &    54595  &  2.09809107745(2)   &   $-$8.7669(1)    & 1.60(3)       & 4363(92)  	  &  52473-56716  &    11.6  &  6054      &563  & 0.08 &   $-$5   & 10153    \\
J0653+8051      &  B0643+80    &    54593  &  0.8234231200677(7) &   $-$2.575859(7)  &  $-$0.018(2)  & $-$197(25)  	  &  52485-56700  &    11.5  &  860      &290  & 0.12 &   $-$3   & 2.866   \\
J0754+3231      &  B0751+32    &    54584  &  0.6933129729262(6) &   $-$0.51932(1)   &  ---          & ---  	          &  52486-56682  &    11.5  &  1233     &303  & 0.08 &   $-$0.6  & 0.495    \\
\\
J0823+0159      &  B0820+02    &    54587  &  1.156239303138(2)  &   $-$0.14009(2)   &  ---          & ---  		  &  52473-56700  &    11.6  &  1491  &403  & 0.25 &   $-$4.5     & 33.74    \\
J0908$-$1739    &  B0906$-$17  &    54601  &  2.489878836633(2)  &   $-$4.14908(3)   &  0.017(7)     & 210(96)  	  &  52485-56716  &    11.6  &  876    &437  & 0.15 &   $-$1.5      & 0.137    \\
J0944$-$1354$\rm ^{I}$ &  B0942$-$13  &    54319  &  1.7535733132312(4) &   $-$0.139213(9)  &  ---          & ---  	   	  &  52473-56166  &    10.1  &  308   &303  & 0.3  &   $-$2       & 0.034    \\
J1047$-$3032    &  ---         &    53571  &  3.02729343359(6)   &   $-$0.557(1)     &  0.8(6)       & 360000(630000) 	  &  52486-54656  &    5.9   &  5138   &141  & 0.1  &   $-$0.1      & 2.367  \\
J1141$-$3322$\rm ^{I}$ &  ---         &    54596  &  3.43091208067(1)   &   $-$5.4710(1)    &  $-$0.84(3)   & $-$8700(240)  	  &  52473-56718  &    11.6  &  1999   &625  & 0.08 &   $-$4       & 1592   \\
\\
J1311$-$1228    &  B1309$-$12  &    54371  &  2.234548091286(2)  &   $-$0.75357(2)   &   0.014(7)    & 3020(2050)  	  &   52472-56270  & 10.4   &  608    &   291  &  0.12  &  $-$2   & 0.166   \\
J1527$-$3931$\rm ^{I}$ &  B1524$-$39  &    54217  &  0.413633385629(6)  &   $-$3.26113(6)   &  0.08(2)      & 276(5)  	  	  &  52500-55934  & 9.4    &  2784   &59   &  0.12  &  $-$2.5    & 78.42   \\
J1543$-$0620    &  B1540$-$06  &    54601  &  1.410308987446(4)  &   $-$1.74562(3)   &  0.07(1)      & 2050(600)  	  &  52485-56717  & 11.6   &  2506   &416  &  0.12  &  $-$5.5    & 244.8  \\
J1555$-$2341    &  B1552$-$23  &    54596  &  1.877659122759(8)  &   $-$2.45793(6)   &  0.15(2)      & 4700(350)  	  &  52473-56718  & 11.6   &  1682   &340  &  0.08  &  $-$3.5    & 246.4   \\
J1603$-$2531$\rm ^{I}$ &  ---         &    54602  &  3.53268367519(3)   &   $-$19.822(3)    &  $-$2.0(1)    & $-$1097(102) 	  &  52486-56718  & 11.6   &12138   &744  &  0.08  &  $-$5.3     & 20153  \\
\\
J1614+0737      &  B1612+07    &    54520  &  0.828636073261(6)  &   $-$1.6208(1)    &  $-$0.08(2)   & $-$808(327) 	  &  52486-56555  &  11.1  &  3261    &134  &  0.1  &   $-$2      & 6.565   \\
J1652$-$2404    &  B1649$-$23  &    54602  &  0.586943754100(3)  &   $-$1.08858(3)   &  ---          & ---		  &  52486-56719  &  11.6  &  4203   &192  &  0.12 &   $-$2      & 28.89   \\
J1700$-$3312$\rm ^{I}$ &  ---         &    54594  &  0.736209749247(2)  &   $-$2.55373(2)   &  0.073(6)     & 807(85)		  &  52486-56702   &   11.5 &  1259  &   211  &  0.12 &   $-$2    & 21.80    \\
J1703$-$1846    &  B1700$-$18  &    54602  &  1.243252566501(4)  &   $-$2.67557(3)   &  $-$0.01(1)   & $-$192(162)	  &  52486-56717  &  11.6  &  1242    &202  &  0.12 &   $-$0.5    & 0.299   \\
J1708$-$3426$\rm ^{I}$ &  ---         &    54565  &  1.444847419693(3)  &   $-$8.77406(2)   &  $-$0.190(7)  & $-$365(13)	  &  52486-56645  &  11.4  &  1034    &303  &  0.12 &   $-$4       & 87.50    \\
\\
J1717$-$3425$\rm ^{I}$ &  B1714$-$34  &    54459  &  1.52368692591(5)   &   $-$2.27479(5)   &  $-$0.2(2)    & 10(44)		  &  52494-56423  &  10.8   &  19214    &273  & 0.2  &   $-$7      & 8768  \\
J1720$-$2933    &  B1717$-$29  &    54550  &  1.611736855430(8)  &   $-$1.93596(6)   &  0.21(2)      & 10790(460)	  &  52494-56605  & 11.3    &  2205     &223  & 0.08 &   $-$4      & 492.5   \\
J1723$-$3659    &  ---         &    54434  &  4.9328092123(1)    &   $-$195.346(2)   &  $-$9.9(4)    & $-$140(7)	  &  52495-56374  &  10.6   &  16535  &255  & 0.16 &   $-$5	       & 19177   \\
J1733$-$2228    &  B1730$-$22  &    54132  &  1.147206212770(4)  &   $-$0.05628(3)   &  $-$0.013(8)  & 176000(445000)	  &  51561-56704  & 14.1     &  3030    &390  &  0.12 &   $-$2   & 4.462   \\
J1735$-$0724    &  B1732$-$07  &    54602  &  2.384725540828(4)  &   $-$6.91513(3)   &  $-$0.081(9)  & $-$407(48)	  &  52487-56717  &    11.6  & 993  &380  &  0.15 &   $-$3.5      & 16.63  \\
\\
J1739$-$2903$\rm ^{I}$ &  B1736$-$29  &  54598    &  3.097068872447(6)  &   $-$75.54400(5)  &  0.13(2)      & 58.8(8)		  &  52495-56702  &    11.5  & 965   &307  &  0.08 &   $-$6   & 5873  \\
J1741$-$0840    &  B1738$-$08  &  54608    &  0.489456220133(4)  &   $-$0.54493(3)   &  0.01(1)      & 2199(1660)	  &  52500-56717  &    11.5  & 4385 &397  &  0.12 &   $-$1     & 2.192  \\
J1743$-$3150$\rm ^{I}$ &  B1740$-$31  &  54599    &  0.4141436626828(2) &   $-$20.71559(2)  & 0.067(5)      & 6.5(7)		  &  52495-56704  &    11.5  &  3053  &222  &  0.08 &   $-$3   & 598.3  \\
J1748$-$1300$\rm ^{I}$ &  B1745$-$12  &  54516    &  2.537209292005(4)  &   $-$7.80772(4)   &  $-$0.62(1)   & $-$2493(43)	  &  52487-56545  &  11.1    &  1041  &355  & 0.08  &  $-$6   & 1221  \\
J1750$-$2438    &  ---         &  54540    &  1.40292420110(2)   &   $-$21.2145(2)   &  $-$0.69(5)   & $-$216(15)	  &  52504-56575  &    11.1  &  5883 &165  & 0.15  &  $-$6     & 3517   \\
\\
J1759$-$1940$\rm ^{I}$ &  ---         &  53575    &   3.92587364700(3)  &   $-$1.441(1)     &   $-$         & $-$73000(117000)   &   52494-54656 &   5.9    &   3795   & 110  & 0.8  & $-$0.1  & 0.696   \\
J1759$-$2922    &  ---         &  54589    &  1.74094171201(1)   &   $-$14.0276(1)   &  0.07(2)      & 33(19)		  &  52496-56683  &    11.5  &  1142 &149  & 0.12  &  $-$1.5   & 7.180      \\
J1807$-$2715$\rm ^{I}$&  B1804$-$27  &  54607  &  1.208048348545(3)  &   $-$17.78871(3)     &  $-$0.096(9)  & $-$64(5)		  &  52496-56719  & 11.6      &  1794 &252  & 0.15  &  $-$5    & 411.5    \\
J1808$-$2057   &  B1805$-$20  &  54599  &  1.088829195129(7)  &   $-$20.23697(7)     & 0.83(2)       & 222(6)		  &  52496-56702  &  11.5     &  44484 &302  & 0.12  &  $-$5   & 5703   \\
J1813+4013     &  B1811+40    &  54618  &  1.074009868890(8)  &   $-$2.93759(6)      & 0.04(2)       & 488(234)		  &  52518-56718  &  11.5     &  2121   &172  & 0.12 &   $-$1  & 4.179       \\
\\
J1817$-$3618$\rm ^{I}$&  B1813$-$36  &  54520  &  2.583853763415(3)  &   $-$13.5984(3)      &  6.1(1)       & 4935(270)	  &  52496-56543  &  11.1     & 10234 &240  & 0.15 &   $-$6.5  & 9662    \\
J1820$-$1346   &  B1817$-$13  &  54607  &  1.085232316695(4)  &   $-$5.28865(3)      &  0.015(9)     & 60(34)		  &  52496-56719  &  11.6     &  1577  &303  & 0.2  &   $-$3   & 21.06      \\
J1822$-$2256   &  B1819$-$22  &  54607  &   0.533541276837(7) &   $-$0.38526(9)      &   0.03(2)     & 5500(6900)	  &   52496-56719 &   11.6    &  2939  &  237  & 0.12 & $-$0.5 & 16.299   \\
J1823$-$3106$\rm ^{I}$&  B1820$-$31  &  54607  &  3.52043890195(2)   &   $-$36.2811(2)      & $-$5.94(6)    & $-$1626(19)	  &  52496-56719  &  11.6    & 3989    &283  & 0.08 &   $-$7  & 238769  \\
J1825+0004     &  B1822+00    &  54606  &  1.283779805919(5)  &   $-$1.44079(5)      &  $-$0.04(2)   & $-$2480(800) 	  &  52494-56718  &    11.6  &  2119  &145  & 0.12 &   $-$1    & 4.179     \\
\\
J1830$-$1135$\rm ^{I}$&  ---         &  54631  &  0.160731253032(6)  &   $-$1.23084(5)      &  0.008(1)     & 835(134)		  &  52578-56684  &    11.2  &  9051  &56   & 0.15 &   $-$6.5  & 9662  \\
J1833$-$0338   &  B1831$-$03  &  54603  &  1.45619318683(2)   &   $-$88.15136(2)     &  1.71(6)      & 34(1)		  &  52487-56719  &  11.6   &  12173 &322  & 0.2  &   $-$3    & 21.06  \\
J1836$-$0436$\rm ^{I}$&  B1834$-$04  &  54515  &  2.822966949325(7)  &   $-$13.21827(8)     & 0.03(2)       & 138(42)		  &  52487-56542  &    11.1  &  1650  &337  & 0.12 &   $-$0.5 & 16.29    \\
J1839$-$0643$\rm ^{I}$&  ---         &  54603  &  2.224450759063(9)  &   $-$17.99075(9)     &  $-$0.24(3)   & $-$230(20)	  &  52487-56719  &    11.6   & 2183 &175 & 0.12 &   $-$2      & 37.89    \\
J1844$-$0433   &  B1841$-$04  &  54523  &  1.009052932303(5)  &   $-$3.98552(5)      &  0.02(1)      & 134(82)		  &  52470-56575  &  11.2    & 2058   &129  &  0.15 &  $-$1.5 &  2.741     \\
\\
J1844+1454$\rm ^{I}$ &  B1842+14    &  54602  &  2.66336954932(4)   &   $-$13.2824(2)       &  $-$4.03(8)   & $-$6122(119)	  &  52487-56717  &    11.6 &  8574  &341  &  0.08 &   $-$5.5  &  80583 \\
J1848$-$1952  &  B1845$-$19  &  54153  &  0.232115465947(2)  &   $-$1.254379(9)      & 0.002(2)      & 42(36)		  &  51588-56718  &  14.1     & 3234   &126  &  0.12 &  $-$0.5 &  0.965     \\
J1850+1335    &  B1848+13    &  54587  &  2.893664785007(5)  &   $-$12.50284(3)      &  $-$0.03(1)   & $-$63(20)	  &  52473-56701  &    11.6   & 723  &519  &  0.12 &   $-$3.5  &  23.43    \\
J1857+0526$\rm ^{I}$ &  ---         &  54619  &  2.857527636028(4)  &   $-$56.58594(4)      & 0.19(1)       & 10(1)		  &  52520-56718  &  11.5   &  743      &230  &  0.15  &  $-$4 & 54.09      \\
J1901+0716$\rm ^{I}$ &  B1859+07    &  54537  &  1.552796029034(6)  &   $-$5.4475(8)        &  $-$8.3(2)    & $-$43740(1090)	  &  52469-56605  &  11.3   &  29067    &307  &  0.08  &  $-$6 & 1170580  \\
\\
J1902+0556    &  B1900+05    &  54288  &  1.339436797083(5)  &   $-$23.09011(5)      & 0.09(2)       & 23(5)		  &  52473-56103  &  9.9    &  2078     &237  & 0.08 & $-$2.5  & 40.64      \\
J1903+0135    &  B1900+01    &  54587  &  1.371166728984(4)  &   $-$7.58499(5)       &  $-$0.02(1)   & $-$27(29)	  &  52473-56700  &    11.6    & 2046   &315  &  0.3 & $-$6.5  & 38.42       \\
J1905+0709    &  B1903+07    &  54219  &  1.54310964912(9)   &   $-$11.8418(9)       &  $-$7.4(3)    & $-$9305(366)	  &  52473-55965  &    9.6    & 19908  &267  &  0.12  & $-$5.5 & 583922   \\
J1909+0007$\rm ^{I}$ &  B1907+00    &  54184  &  0.983331573789(6)  &   $-$5.33435(8)       &  $-$0.88(2)   & $-$3047(87)	  &  52470-55898  &    9.4     & 2455   &159  & 0.12 &   $-$6  & 2308     \\
\\
J1909+1102    &  B1907+10    &  55123  &  3.52557660509(4)   &   $-$32.2862(6)       &  $-$2.1(3)    & $-$680(89)	  &  53699-56546  &    7.8    & 3986  &165  & 0.11 &   $-$5.5   & 9627    \\
J1912+2104    &  B1910+20    &  54529  &  0.447833217497(2)  &   $-$2.04076(3)       &  0.053(6)     & 572(65)		  &  52515-56542  &    11.0   &  2003  &146  & 0.12 &   $-$1.5 & 18.45        \\
J1916+0951    &  B1914+09    &  54587  &  3.700202742932(8)  &   $-$34.48575(6)      & 0.10(2)       & 23(6)		  &  52473-56700  &    11.6  &  922    &249  & 0.2   &  $-$4.5 & 31.14        \\
J1918+1444$\rm ^{I}$ &  B1916+14    &  54644  &  0.84665819571(2)   &   $-$152.0376(3)      & 6.39(8)       & 22.9(3)		  &  52569-56719  &    11.4  &  18531 &206  & 0.08  &  $-$6    & 3908260  \\
J1921+1419$\rm ^{I}$ &  B1919+14    &  54603  &  1.61763759009(1)   &   $-$14.65135(7)      &  0.53(2)      & 398(16)		  &  52487-56718  &    11.6  &  2632  &241  & 0.08  &  $-$4.5  & 3011     \\
\\
J1926+0431    &  B1923+04    &  54606  &  0.931029831258(3)  &   $-$2.13009(3)       &  $-$0.088(8)  & $-$1810(170)	  &  52497-56716  &    11.6  &  1261   &237  & 0.1   &  $-$3.5 & 208.9      \\
J1932+2020$\rm ^{I}$& B1929+20   &   54596  &  3.72829493293(2)   &  $-$58.66093(2)         &   $-$0.09(6)  & $-$10(6)		  & 52472-56720   &   11.6   &  3919  &   285  & 0.15  &  $-$5  & 677.4       \\
J1937+2544   & B1935+25   &   54508  &  4.975607766809(5)  &  $-$15.92257(6)         &  0.09(1)      & $-$83(18)	  &  52473-56543  & 11.1   & 606  & 281       & 0.15  &  $-$2.5  & 0.228         \\
J1943$-$1237 & B1940$-$12 &   54561  &  1.028351970679(4)  &  $-$1.75106(4)          &  $-$0.46(1)   & $-$15430(420)	  &  52497-56624  & 11.3   & 1487 & 218        & 0.08  &  $-$5.5 & 5691      \\
J2004+3137$\rm ^{I}$& B2002+31   &   54539  &  0.473642777972(2)  &   $-$16.71669(2)        & 0.338(6)      & 62.6(8)		  &  52488-56590  &    11.2  &  2452    &245  & 0.08 &   $-$6.5  & 58155     \\
\\
J2029+3744   & B2027+37   &   54150  &  0.82182139886(3)   &   $-$8.36473(3)         &  $-$0.3(2)    & $-$173(121)	  &  52474-55826  &    9.2   &  14615   &215  & 0.12 &   $-$6    & 26774   \\
J2046$-$0421 & B2043$-$04 &   54607  &  0.646437948182(2)  &   $-$0.61485(2)         &  0.012(6)     & 2100(1080)	  &  52497-56717  &  11.6    &  1748    &214  & 0.12 &   $-$0.5  & 0.410       \\
J2055+2209   & B2053+21   &   54546  &  1.226720413993(2)  &   $-$2.01667(2)         &  $-$0.005(5)  & 78(129)		  & 56605-54542   &   11.3   &  644   &  193  & 0.15 &   $-$1    & 0.114       \\
J2113+2754   & B2110+27   &   54595  &  0.8313567139474(5) &   $-$1.812815(5)        &  $-$0.028(1)  & $-$622(60)	  &  52474-56717  &    11.6  & 939      &255  & 0.1  &   $-$2    & 1.847       \\
J2149+6329$\rm ^{I}$& B2148+63   &   54595  &  2.630606301836(2)  &   $-$1.17881(1)         &  $-$0.181(6)  & $-$33800(900)      &  52474-56717  &    11.6   &  614    &441  & 0.12 &  $-$4.5   & 112.1    \\
\\
J2212+2933   & B2210+29   &   54588  &  0.995428242776(3)  &   $-$0.48995(3)         &  0.03(1)      & 14550(300)        &  52474-56703  &    11.6   &  1372   &178  & 0.12 &  $-$2     & 2.690  \\
J2229+6205$\rm ^{I}$& B2227+61   &   54587  &  2.257053611029(7)  &   $-$11.49531(5)        &  0.49(2)      & 849(24)		  &  52470-56703  &    11.6   &  1235   &406  & 0.08 &  $-$4.5   & 2091    \\
J2305+3100$\rm ^{I}$& B2303+30   &   54602  &  0.6345629290550(8) &  $-$1.164933(8)         & 0.0131(2)     & 560(144)		  &   52488-56717 &   11.6  &  1082   & 277  & 0.15 &  $-$3.5   & 3.484       \\
J2317+2149$\rm ^{I}$& B2315+21   &   54606  &  0.6922074330766(4) &  $-$0.501725(10)        &   ---         & ---		  &   52493-56718 &   11.6   &  736    & 184  & 0.15 &  $-$0.5   & 0.251       \\
J2325+6316   & B2323+63   &   54594  &  0.696228000638(7)  &   $-$1.36825(5)         &  0.05(2)      & $-$1690(590)	  &  52471-56718  &    11.6  &  5953   &400  & 0.1  &  $-$1.5   & 21.34      \\
\enddata
\end{deluxetable*}
\end{longrotatetable}

\subsection{ Glitches in three pulsars}\label{sec:glitch}
\begin{figure*}
\begin{center}
\centering
\includegraphics[width=2.3in,height=2.2in,angle=0]{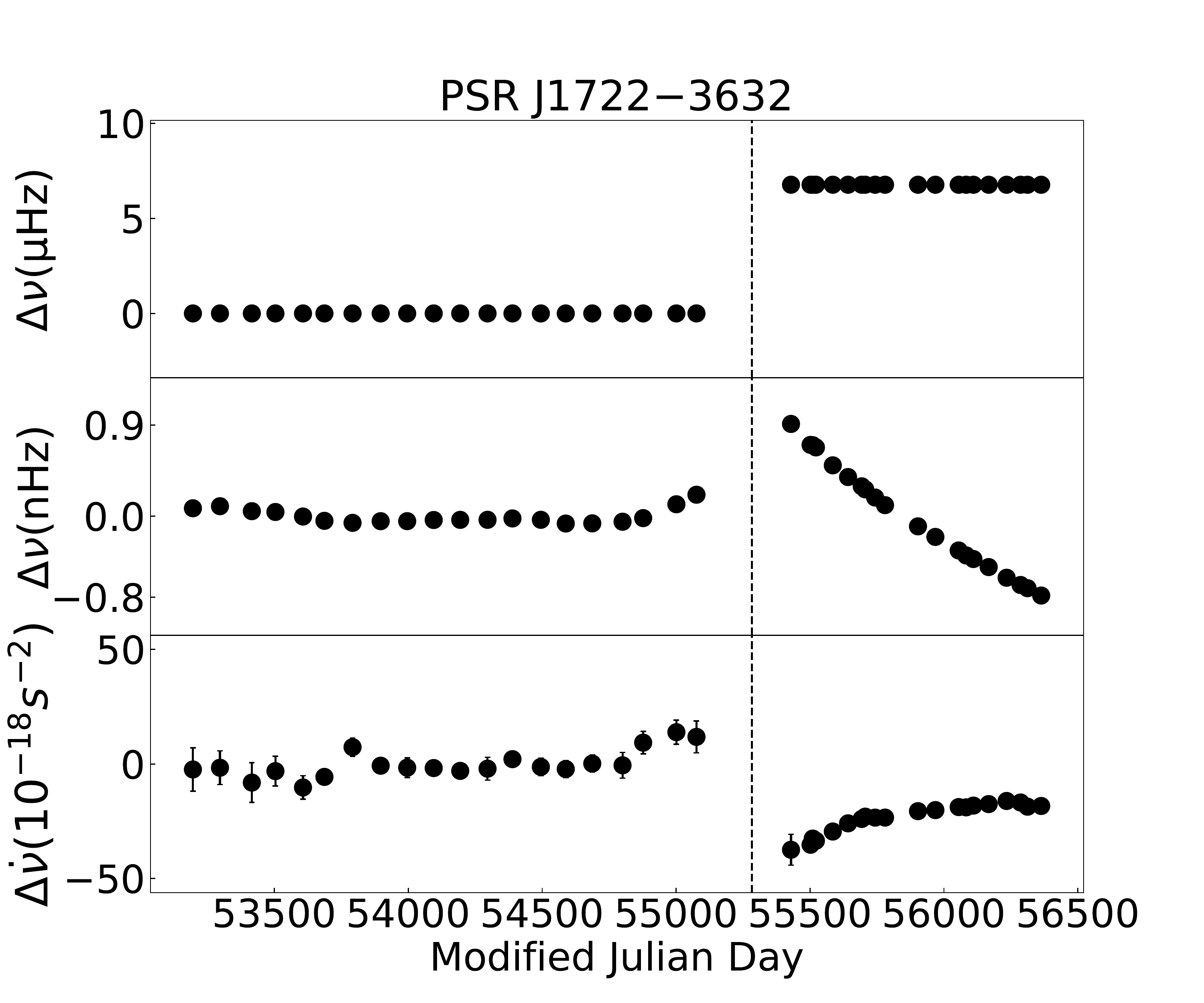}
\includegraphics[width=2.3in,height=2.2in,angle=0]{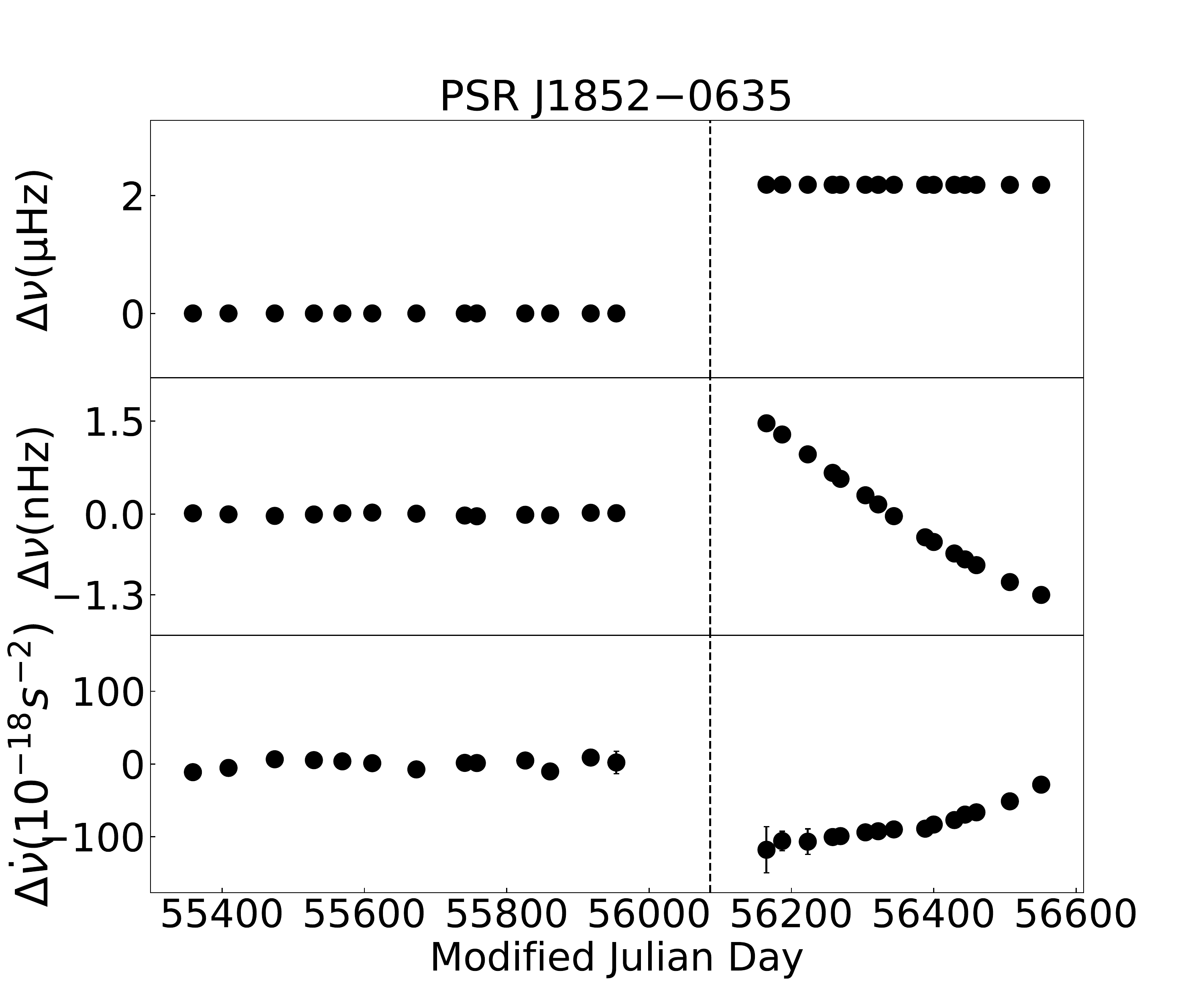}
\includegraphics[width=2.3in,height=2.2in,angle=0]{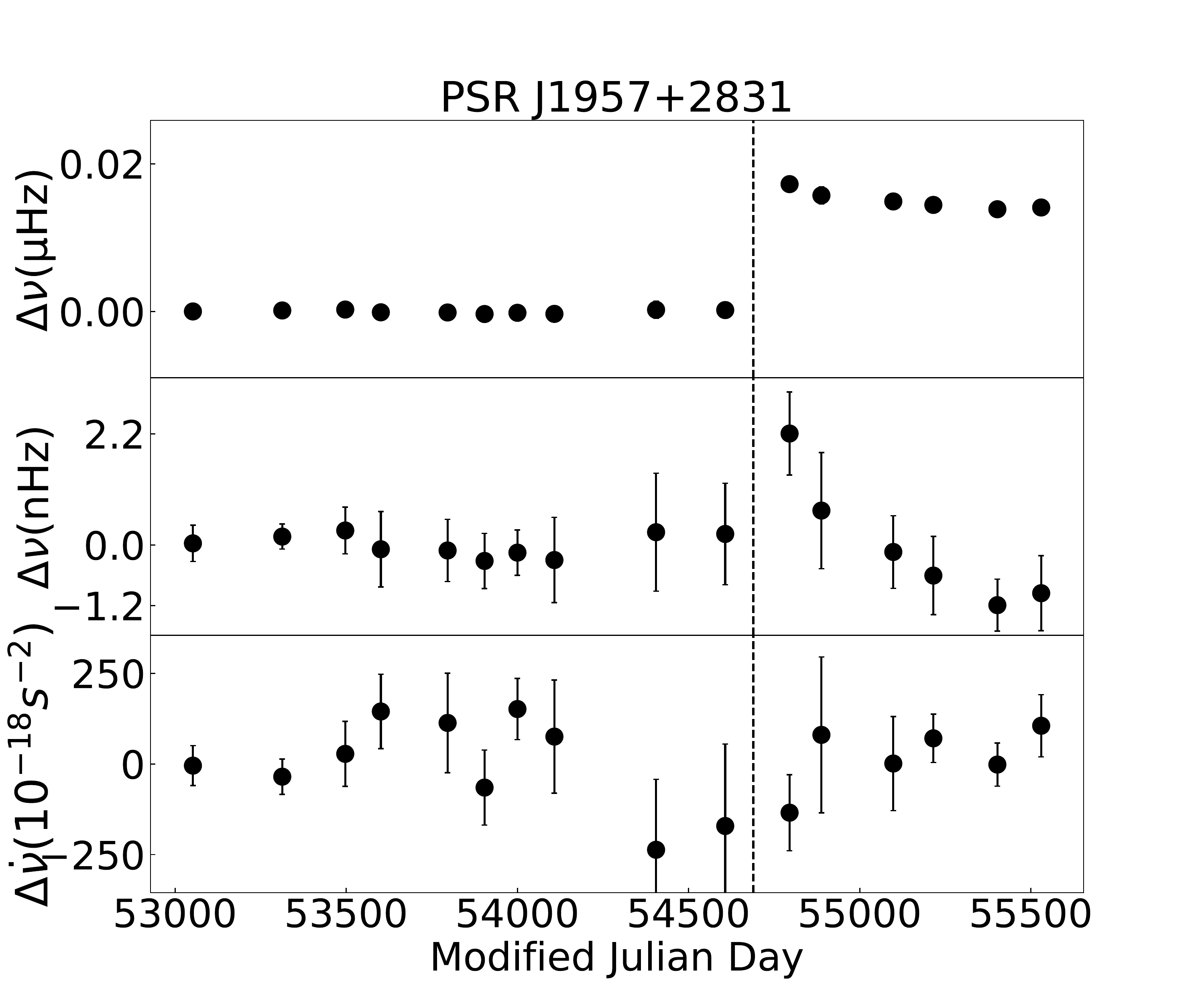}
\caption{The glitches of PSRs J1722$-$3632, J1852$-$0635 and J1957+2831. For each pulsar, the top panel shows the spin-frequency residuals $\Delta{\nu}$ relative to the pre-glitch timing solution; the middle panel is an expanded plot of $\Delta{\nu}$ where the mean post-glitch value is subtracted from the post-glitch values; the bottom panel shows the variations of $\dot{\nu}$ relative to the mean pre-glitch value. The dashed line in each panel indicates the glitch epoch.}
\label{fg:glitches_mu}
\end{center}
\end{figure*}

Glitches were detected in three pulsars: PSRs J1722$-$3632, J1852$-$0635 and J1957+2831. For pulsars that have undergone large glitches, it is difficult to determine the timing solution with the whole data span and so we have separately fitted the pre- and post-glitch ToAs. The midpoint of the last pre-glitch observation  ($\rm T_1$) and the first post-glitch observation  ($\rm T_2$) is adopted as the glitch epoch. The uncertainty of glitch epoch was estimated as $\rm (T_{2} -T_{1} )/4$, which corresponds to $1 \sigma $ of a square distribution between $\rm T_{1}$ and $\rm T_{2}$ \citep[see:][]{els+11}. The evolution of spin frequency $\nu$ and frequency derivative $\dot\nu$ for the three glitch pulsars is shown in Figure~\ref{fg:glitches_mu}. Values of $\nu$  and $\dot\nu$  were derived from independent fits to short sections of data (100~d -- 200~d), each of which contains 12 -- 20 ToAs. The plotted values were then obtained by subtracting the (extrapolated) pre-glitch model.

\begin{figure}
\begin{center}
\centering
\includegraphics[width=3.5in,height=4in,angle=0]{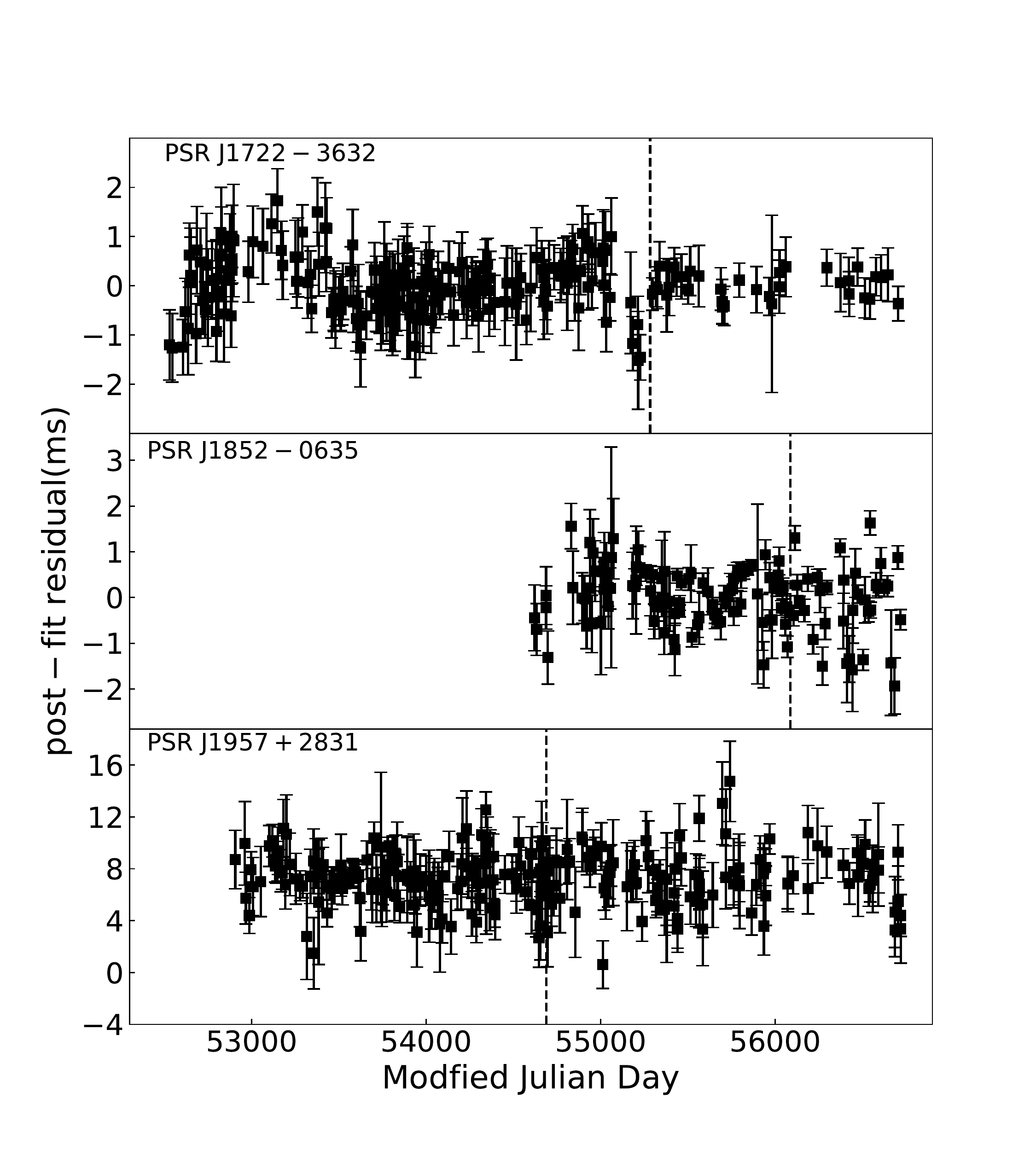}
\caption{Timing residuals of PSRs J1722$-$3632, J1852$-$0635 and J1957+2831 after fitting for all glitch parameters. The vertical black dashed line in each panel  indicates the glitch epoch.}
\label{fg:residuals_glt_psr}
\end{center}
\end{figure}

Table~\ref{tb:glitch_tim} gives the timing solutions for these three pulsars.
We also fitted the glitch parameters using the model:
\begin{equation}\label{eq:gltphi}
  \phi_g=\Delta\phi+\Delta\nu_p(t-t_g)+\frac{1}{2}
  \Delta\dot\nu_p(t-t_g)^2+[1-e^{-(t-t_g)/\tau_d}]\Delta\nu_d\tau_d
\end{equation}
where $t_g$ is the glitch epoch and $\Delta\phi$ is an increment of pulse phase between the pre- and post-glitch data. The glitch event is characterised by permanent increments in the spin frequency $\Delta\nu_p$ and first frequency derivative $\Delta\dot\nu_p$ and a transient frequency increment $\Delta\nu_d$ which decays exponentially with a time scale $\tau_d$. Then, the increment in the spin frequency and the frequency first derivative at the glitch are described by $\Delta\nu_g=\Delta\nu_p+\Delta\nu_d$ and $\Delta\dot\nu_g=\Delta\dot\nu_p-\frac{\Delta\nu_d}{\tau_d}$. The glitch recovery factor is defined by $Q \equiv \Delta\nu_d/\Delta\nu_g$ \citep[see:][]{ymw+10,ymh+13}. Table~\ref{tb:glitches_table} gives the glitch parameters, where the uncertainties are 1$\sigma$ from the error propagation formula. Figure~\ref{fg:residuals_glt_psr} is the timing residuals of three glitch pulsars after fitting all of the glitch parameters.

\begin{sidewaystable}[htbp]
\centering
\footnotesize
\caption{Timing parameters for PSRs J1722$-$3632, J1852$-$0635 and J1957+2831. The uncertainties in parentheses are in the last quoted digit and are $1\sigma$ as given by \textsc{Tempo2}.}
\label{tb:glitch_tim}
\begin{tabular}{lllllll}
\hline
\hline
\multicolumn{1}{l}{PSR}  & \multicolumn{2}{c}{J1722$-$3632} &     \multicolumn{2}{c}{J1852$-$0635}   &     \multicolumn{2}{c}{J1957+2831} \\
Parameters                       & Pre-glitch                       &  Post-glitch             &       Pre-glitch      &  Post-glitch  &       Pre-glitch      &  Post-glitch \\
\hline
RA  (h m s)                      & 17:22:09.793(2)                  &  17:22:09.798(2)         &  18:52:57.461(8)$^*$      &  18:52:57.461(8)$^*$  &19:57:19.397(8)$^*$ & 19:57:19.397(8)$^*$ \\	
DEC (d m s)                      & $-$36:32:55.0(1)                 &  $-$36:32:55.2(1)        &  $-$06:36:00.4(4)     &  $-$06:36:00.4(4)  & +28:31:43.8(1) & +28:31:43.8(1) \\
$\rm \mu_{\alpha} (mas~ yr^{-1})$                   & 12(14)                           & 12(14)                   &  ---                   & ---               &  ---                   & --- \\
$\rm \mu_{\delta } (mas~ yr^{-1})$                   & $-$34(62)                        & $-$34(62)                &  ---                   & ---               &  ---                   & --- \\
$\nu (s^{-1})$                   & 2.505101132060(5)                &  2.50510275676(2)        &  1.907834375341(9)    &  1.90783178192(5)  & 3.2500929512(2) & 3.25008756591(3) \\
$\dot\nu (s^{-2})$  & $ -2.801283(6)\times 10^{-14}$&  $-2.80328(5)\times 10^{-14} $ & $ -5.32256(3)\times 10^{-14}$ & $-5.3309(4)\times 10^{-14} $  & $-3.28527(5)\times 10^{-14}$ & $-3.28597(5)\times 10^{-14}$ \\
$\ddot\nu (s^{-3})$ &  $1.3(3)\times 10^{-26} $     & $ 7(4)\times 10^{-26}	$       &  $1.5(2)\times 10^{-25}$     &  $6(8)\times 10^{-25}$	  & $3(3)\times 10^{-26}$ & $4(3)\times 10^{-26}$ \\
Epoch of period (MJD)                    & 53882                           &  55999                    &  55345                &  56383  & 53803  & 55706 \\	
Epoch of position (MJD)                  & 53882                           &  55999                    &  55345                &  56383  & 53803 & 55706 \\	
Data span (MJD)                       & 52494-55270                     &  55295-56719             &  54620-56069           &  56102-56663   & 52911-54683 & 54694-56719 \\ 	
\multicolumn{1}{l}{ Time units}                            & \multicolumn{6}{c}{ TCB }  \\
\multicolumn{1}{l}{ Solar-system ephemeris model  }        & \multicolumn{6}{c}{ DE421 }    \\
\hline
\end{tabular}
\begin{flushleft}
$^*$ Positions from the ATNF Pulsar Catalogue, V1.61.
\end{flushleft}
\end{sidewaystable}

\begin{table}
\centering
\footnotesize
\caption{Glitch parameters for PSRs J1722$-$3632, J1852$-$0635 and J1957+2831. Uncertainties are from the error propagation formula and are 1$\sigma$.}
\label{tb:glitches_table}
\begin{tabular}{llllllll}
\hline
\hline
\multicolumn{1}{l}{PSR J2000 Name}  & \multicolumn{1}{l}{J1722$-$3632}  &  \multicolumn{1}{l}{J1852$-$0635}   &     \multicolumn{1}{l}{J1957+2831} \\
\hline
Glitch epoch(MJD) &55283(3)& 56086(8) & 54689(3) \\
$\Delta{\nu}_g$ $\rm (10^{-9}Hz)$ & 6769.2(1)& 2182.8(7) & 20.9(3) \\
$\Delta{\nu}_g/{\nu}$ $\rm (10^{-9})$ &2702.18(4) & 1144.1(3) & 6.4(6)  \\
$\Delta\dot{\nu}_g$ $\rm (10^{-17} s^{-2})$ & $-$4.7(5) & $-$15(2) & $-$2.3(4) \\
$\Delta\dot{\nu}_g/\dot{\nu}$ $\rm (10^{-3})$ &0.89(9) & 5.5(6) & 0.7(6) \\
$\Delta{\nu}_d $ $ \rm (10^{-9}Hz)$ & 0.4(1) & 4.7(5) & 4(1)\\
$\tau_d $ $ \rm (d)$ & 240(20) & 400(40)   &  110(10) \\
  Q & 0.00006(2)  & 0.0022(2)  & 0.2(3) \\
Data range (MJD) & 53034-56879 & 54619-56718 & 52911-56719 \\
\hline
\end{tabular}
\end{table}

\subsubsection{PSR J1722$-$3632}
PSR J1722$-$3632 (B1718$-$36) is a relatively old pulsar (characteristic age $\tau_c \sim 1.4\times 10^6$~yr) discovered in a Parkes survey of the southern Galactic plane by \citet{jlm+92}. No previous glitches have been reported for this pulsar. Nanshan timing observations of PSR J1722$-$3632 commenced in 2002, September. Here, we report the discovery of a large glitch with $\Delta\nu_g/{\nu} =2702.18(4) \times10^{-9} $ and $\Delta\dot\nu_g/{\dot\nu} = 0.89(9) \times10^{-3}$ that occurred around MJD 55283. As shown in Figure~\ref{fg:glitches_mu}, the jump in $\nu$ decays only slightly although there is indication of an exponential recovery in both $\nu$ and $\dot\nu$. From the fit of the glitch model (Equation~\ref{eq:gltphi}) we derive a fractional decay $Q$ of just 0.00006(2) (Table~\ref{tb:glitches_table}). This is consistent with earlier results that small values of Q are commonly seen for large glitches \citep{ymh+13}. The fitted value for the time constant $\tau_d$ of the exponential decay is about 240 days (Table~\ref{tb:glitches_table}). Another striking point about the post-glitch recovery in this pulsar is that, for the available data span, only about half of the increment in $\dot\nu$ at the glitch, ($\Delta\dot\nu_g$) recovers exponentially. It is possible that future observations will show a linear increase in $\dot\nu$ as is commonly observed after the exponential recovery in large glitches \citep{ymh+13}.

\subsubsection{PSR J1852$-$0635} This pulsar has a characteristic age $\tau_c \sim 5.7\times 10^5$~yr and no glitch event has previously been reported for it. Nanshan observations of PSR J1852$-$0635 started in 2008, June. As shown in Figure~\ref{fg:glitches_mu}, we detected a relatively large glitch near MJD 56086 (2012, June) with a fractional frequency change $\Delta{\nu_g}/{\nu} = 1144.1(3) \times 10^{-9}$. Again, the fractional post-glitch decay is very small, with $Q$ about 0.0022(2) (Table~\ref{tb:glitches_table}). There is some indication of an exponential recovery in $\nu$ with time-scale about 400~d, but the recovery in $\dot\nu$ appears more linear than exponential.

\subsubsection{PSR J1957+2831} PSR J1957+2831 was discovered by \citet{llc98} and is associated with the supernova remnant G65.1+0.6 \citep{tl06}. Three small glitches in this pulsar were reported by \citet{els+11}. In our data span (from MJD 52911 to 56719), we detected the third glitch at MJD $\sim 54689$(3) with a fractional frequency change of $\Delta\nu_g/{\nu} = 6.4(6) \times 10^{-9} $. Figure~\ref{fg:glitches_mu} shows a clear  exponential recovery following the glitch which was not detected by \citet{els+11}. We have fitted the glitch parameters using \textsc{Tempo2} and the results are given in Table~\ref{tb:glitches_table}. The exponential decay has a fraction $Q \sim 0.2$ and time-scale $\tau_d \sim 110$~d.

\subsubsection{Other claimed glitches}
\begin{figure}
\begin{center}
\centering
\includegraphics[width=3.5in,height=3.5in,angle=0]{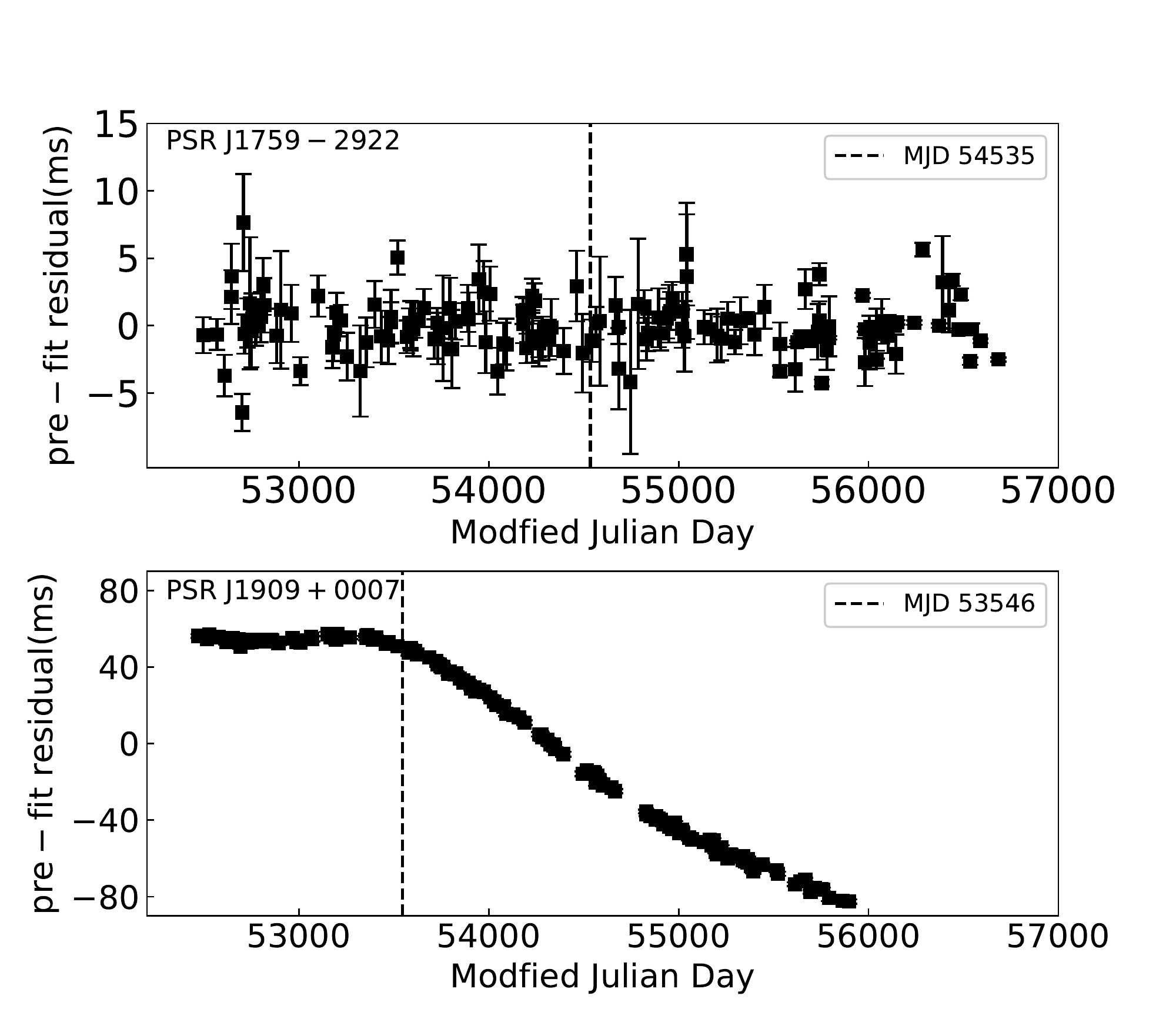}
\caption{Pre-fit timing residuals of PSRs J1759$-$2922 and J1909+0007.}
\label{fg:residuals_1759_1901}
\end{center}
\end{figure}

The Jodrell Bank glitch catalogue reports a small glitch for each of PSRs J1759$-$2922 and J1909+0007 around MJD 54535 and 53546, respectively \citep[the latter also published by][]{els+11}. These reported glitches are within our data spans and the pre-fit timing residuals obtained from our observations are shown in Figure~\ref{fg:residuals_1759_1901}. Our data, which are well sampled, show no evidence for a glitch in either pulsar at the relevant epoch. For PSR J1759$-$2922, the timing residuals are almost white. For PSR J1909+0007 there is a change of pulse frequency around the epoch of the claimed glitch, but it is slow and can be attributed to red timing noise.

\section{Discussion}\label{sec:discussion}
\subsection{Velocities}\label{sbec:velocity}
Although the different estimates of pulsar distance have an impact on the computed transverse velocities (Tables~3 and~\ref{tab:1D_velocities}), the uncertainties of the total proper motions are sufficiently large, so that the computed transverse velocities for a given pulsar are generally consistent within the uncertainties. For example, PSR J1543$-$0620 has transverse velocities, $\rm V^{T}_{Y}$ = 142(78), $\rm V^{T}_{CL}$ = 91(50) and $\rm V^{T}_{D}$ = 400(200) km~s$^{-1}$, which overlap at the 1.5-$\sigma$ level despite a factor of four difference in the estimated distances.

\begin{figure}
\begin{center}
\centering
\includegraphics[width=3.5in,height=3in,angle=0]{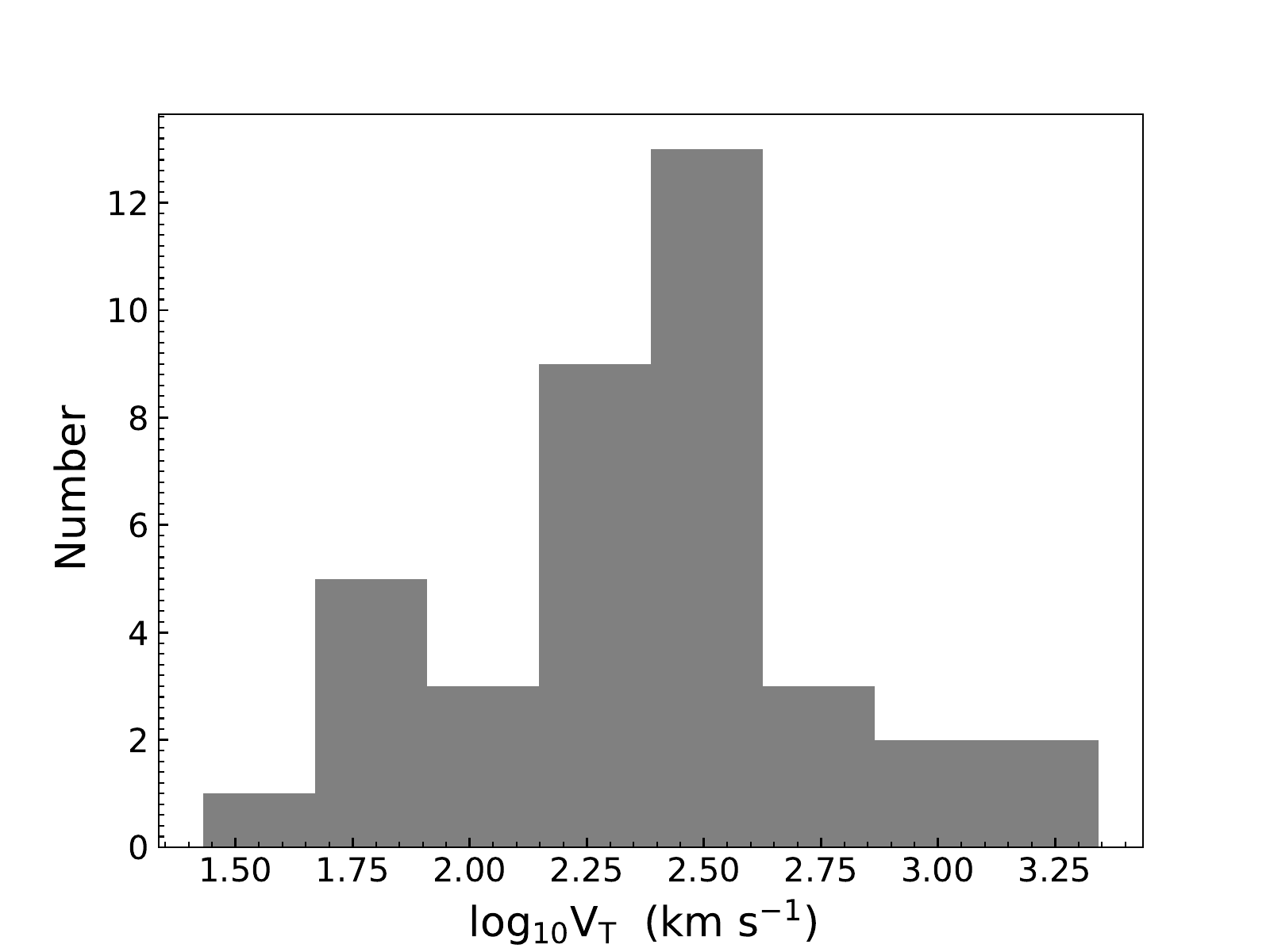}
\caption{Distribution of pulsar transverse velocities, $\rm V_{T}$, for the pulsars in our sample.}
\label{fg:VT_distribution}
\end{center}
\end{figure}

Figure~\ref{fg:VT_distribution} shows the distribution of derived  transverse velocities from our results. For this plot, we show only pulsars with velocity uncertainties less than three times the mean transverse velocity of 270~km~s$^{-1}$ from \citet{hllk05} and we chose the values with the least uncertainty between results from the timing method and the position comparison method. Except for nine pulsars which have a VLBI distance and PSR J0134+2937 and J1311$-$1228 for which the NE2001 distance was adopted, we used distances from the YMW16 model. The mean and rms velocities of the sample are 255 and 168 km $\rm s^{-1}$, which are consistent with previous estimates \citep[e.g.,][]{zhw+05,hllk05}.

\subsection{Glitches}
We have detected glitches in PSRs J1722$-$3632, J1852$-$0635 and J1957+2831. For PSRs J1722$-$3632 and J1852$-$0635, these is the first detected glitches. These pulsars have relatively large characteristic ages among the sample of glitching pulsars, with  $\tau_c \sim$~1.4, 0.6 and 1.6~Myr, respectively. Two of the glitches, in PSRs J1722$-$3632 and J1852$-$0635, are large with $\Delta\nu_g/\nu$ greater than $10^{-6}$ in both cases. The post-glitch recoveries seen in these pulsars are typical of those seen after large glitches. Such exponential recoveries can be interpreted by the "vortex creep" model \citep[see, e.g.,][]{accp93,lsg00} where the rotation of the superfluid interior and the crust return to equilibrium after a glitch.

\begin{figure}
\begin{center}
\centering
\includegraphics[width=3.5in,height=3in,angle=0]{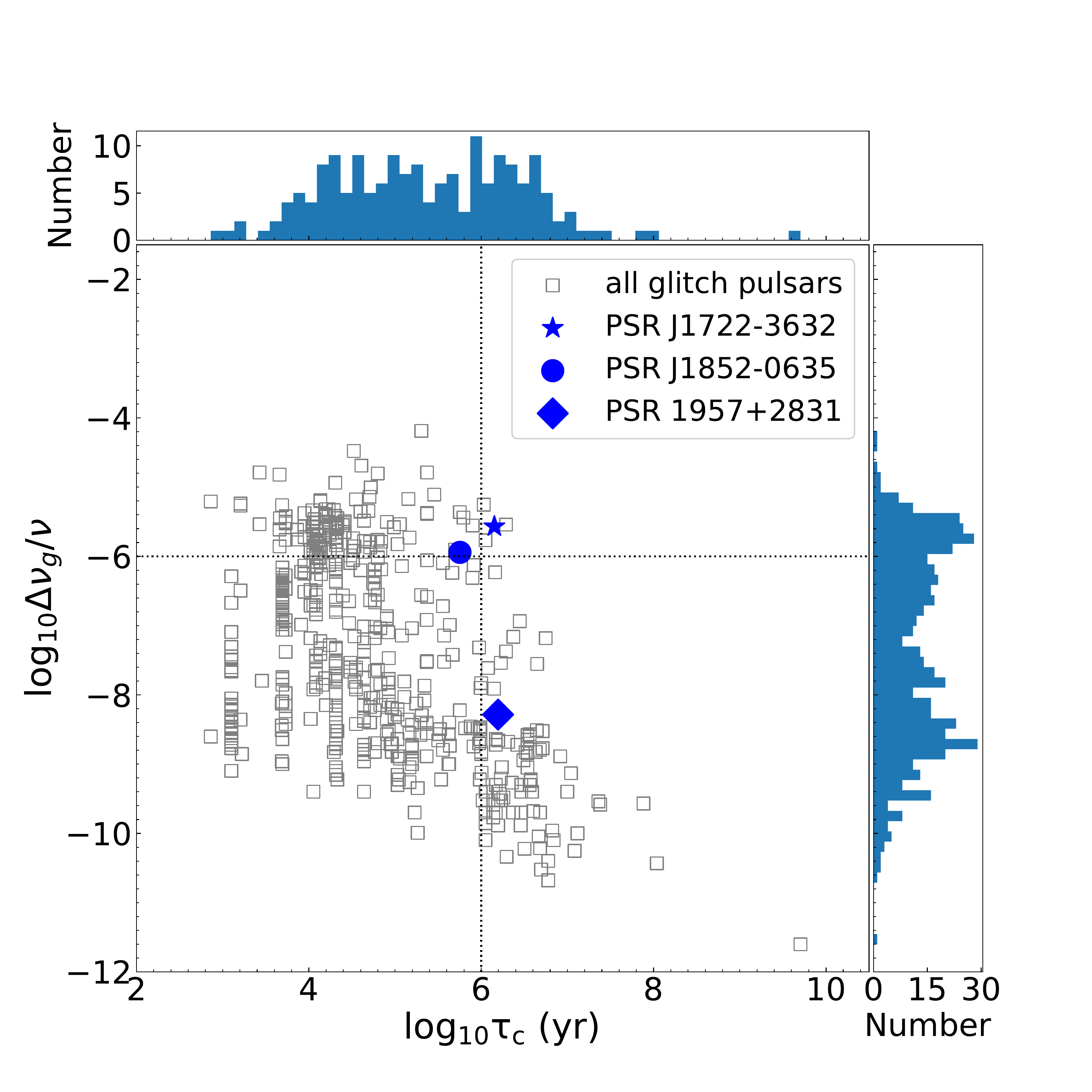}
\caption{Glitch fractional frequency increment $\Delta\nu_g/\nu$ versus characteristic age $\tau_c$ for all pulsars with observed glitches. The blue "$\star$", "$\bullet$" and "$\Diamond$" represent the pulsars reported here, PSRs J1722$-$3632, J1852$-$0635 and J1957+2831, respectively. The vertical dotted line is at $\tau_c =10^6$~yr and the horizontal dotted line is at $\Delta\nu_g/\nu =10^{-6}$. The top and left panels are the distributions of $\tau_c$ and $\Delta\nu_g/\nu$, respectively.}
\label{fg:DF0_F0_vs_age}
\end{center}
\end{figure}

We have investigated the relationship between $\Delta\nu_g/\nu$ and $\tau_c$ for all known glitches as shown in Figure ~\ref{fg:DF0_F0_vs_age}. These data are from the ATNF Glitch Catalogue and the Jodrell Bank Glitch Catalogue and include the pulsars discussed in this paper. As is well known \citep{els+11,ymh+13,FERS+17}, the distribution of relative glitch sizes is bi-modal with a group of "large" glitches with $\Delta\nu_g/\nu \sim 10^{-6}$ and a second broad group of "small" glitches with $\Delta\nu_g/\nu \la 10^{-7}$. As is illustrated in Figure ~\ref{fg:DF0_F0_vs_age}, large glitches have $\Delta\nu_g/\nu$ that is essentially independent of $\tau_c$ but an occurrence rate that is strongly dependent on $\tau_c$, with fewer glitches observed at larger $\tau_c$. Small glitches, on the other hand, have an inverse dependence of $\Delta\nu_g/\nu$ on $\tau_c$ and an occurrence rate that is only weakly dependent of $\tau_c$, at least up to $\tau_c\sim 10^7$~ yr. Almost certainly, the physical mechanisms for these two types of glitches are different. Specifically, few large glitches are detected in old pulsars, with just four pulsars with $\tau_c > 10^6$~yr having large glitches. PSR J1722$-$3632 is the second oldest pulsar known to have a glitch with $\Delta\nu_g/\nu > 10^{-6}$.

\begin{figure}
\begin{center}
\centering
\includegraphics[width=3.5in,height=3in,angle=0]{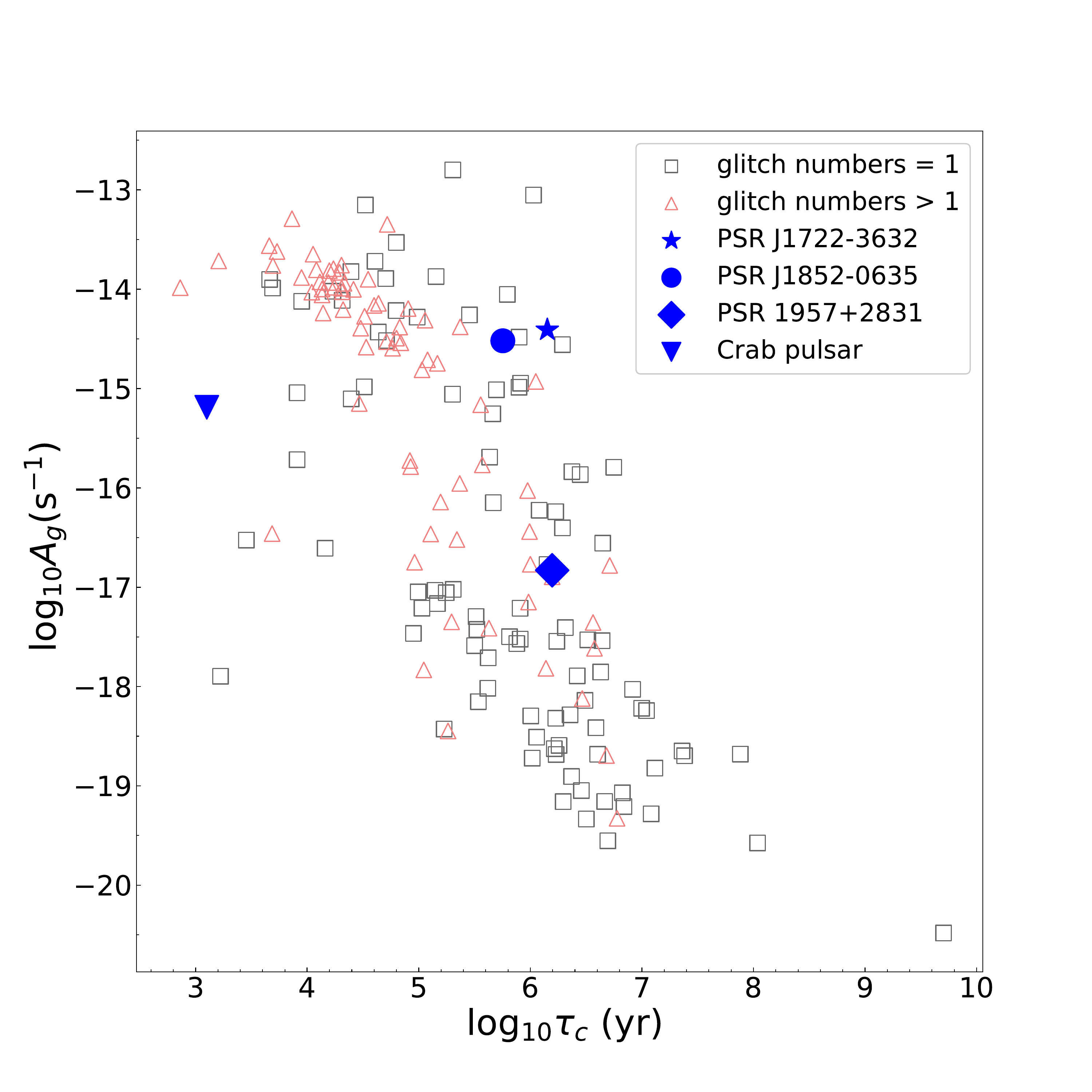}
\caption{Glitch activity parameter $A_g$ versus characteristic age $\tau_c$. The gray "$\square$" and red "$\triangle$" stand for pulsars with one or  more observed glitches, respectively. The blue "$\star$", "$\bullet$", "$\Diamond$" and "$\triangledown$" represent PSRs J1722$-$3632, J1852$-$0635, J1957+2831 and Crab pulsar, respectively.}
\label{fg:glitch_active_parameters_vs_age}
\end{center}
\end{figure}

The glitch activity parameter, generally defined by
\begin{equation}
    A_g \equiv \frac{1}{T}\sum \frac{\Delta \nu_g}{\nu}
\end{equation} \citep{ml90,lsg00}, where $T$ is the total data span over which the pulsar has been observed and $\sum \Delta \nu_g/\nu$ is the sum of the relative glitch frequency increments for all glitches in a given pulsar, is a measure of the impact of glitch activity on the mean slow-down rate. Figure~\ref{fg:glitch_active_parameters_vs_age} shows $A_g$ versus $\tau_c$ for glitching pulsars obtained from the ATNF Glitch Catalogue and the Jodrell Bank Glitch Catalogue as well as the three pulsars discussed here. In the calculation, we have taken the date of pulsar discovery as the start of the data span. For our three glitch pulsars, $T$ extends to the end of our data span. For other pulsars, 2019 January is assumed.

 Figure~\ref{fg:glitch_active_parameters_vs_age} shows that $A_g$ is largest for pulsars with $\tau_c \sim 10^5$~yr, decreasing on average for both younger and older pulsars. Although frequent, glitches tend to be small in very young pulsars such as the Crab pulsar, giving them a low $A_{g}$, as well as very old pulsars (Figure~\ref{fg:DF0_F0_vs_age}). Glitch activity parameters for PSRs J1722$-$3632, J1852$-$0635 and J1957+2831 are $\sim 3.9 \times 10^{-15}$~s$^{-1}$, $ 3.0 \times 10^{-15}$~s$^{-1}$ and $1.5 \times 10^{-17}$~s$^{-1}$, respectively. Since only one glitch has been observed in the 22 and 12-year observational data spans for PSRs J1722$-$3632 and J1852$-$0635, the quoted values are upper limits. Never-the-less, these values are within the observed range of activity parameters for middle-aged pulsars \citep{rgl12}.\footnote{Note that \citet{rgl12} define $A_g \equiv \frac{1}{T}\sum\Delta\nu_g$, that is, the mean increment in $\dot\nu$ due to glitches.}

For the widely accepted model where glitches result from a sudden transfer of angular momentum from an interior superfluid to the neutron-star crust, \citet{lel99} showed that the ratio of the moment of inertia of the interior superfluid, $I_s$, to the moment of inertia of the crust (and any interior components tightly coupled to it), $I_c$, must obey the constraint:
\begin{equation}
\frac{I_{s}}{I_c} \geq \mathcal{G} \equiv 2 \tau_c A_g =  \frac{\bar{\nu}}{|\dot\nu|} \frac{1}{T} \sum \frac{\Delta \nu_g}{\nu}.
\end{equation}
where $\bar{\nu}$ and $|\dot\nu|$ are the average spin rate and spin-down rate of the crust over the period of observations, respectively. $\mathcal{G}$ is an observational parameter, but is known as the "coupling parameter" because of this relation. For the Vela pulsar, $\mathcal{G}$ is about $1.6\%$ and similar values are found for other young and frequently glitching pulsars \citep{heaa15}. However, \citet{aghe12} and \citet{c13} have argued that, because of "entrainment" of superfluid in the inner crust, \citet{lel99} underestimated the value of $I_s/I_c$ by about a factor of four. When the entrainment is considered, using the latest data which contain 20 glitches for the Vela pulsar during $\sim 50$ years of observation, the corresponding $I_{s}/I_{c}$ is about 6.9\%. This leads to the conclusion that the crustal superfluid, by itself, cannot account for large glitches such as those in the Vela pulsar, and hence that another superfluid component, perhaps in the neutron-star core, must contribute to the observed glitches \citep[see also][]{dcg+16,bcn+18}.

\begin{figure*}
\begin{center}
\centering
\includegraphics[width=3.5in,height=3in,angle=0]{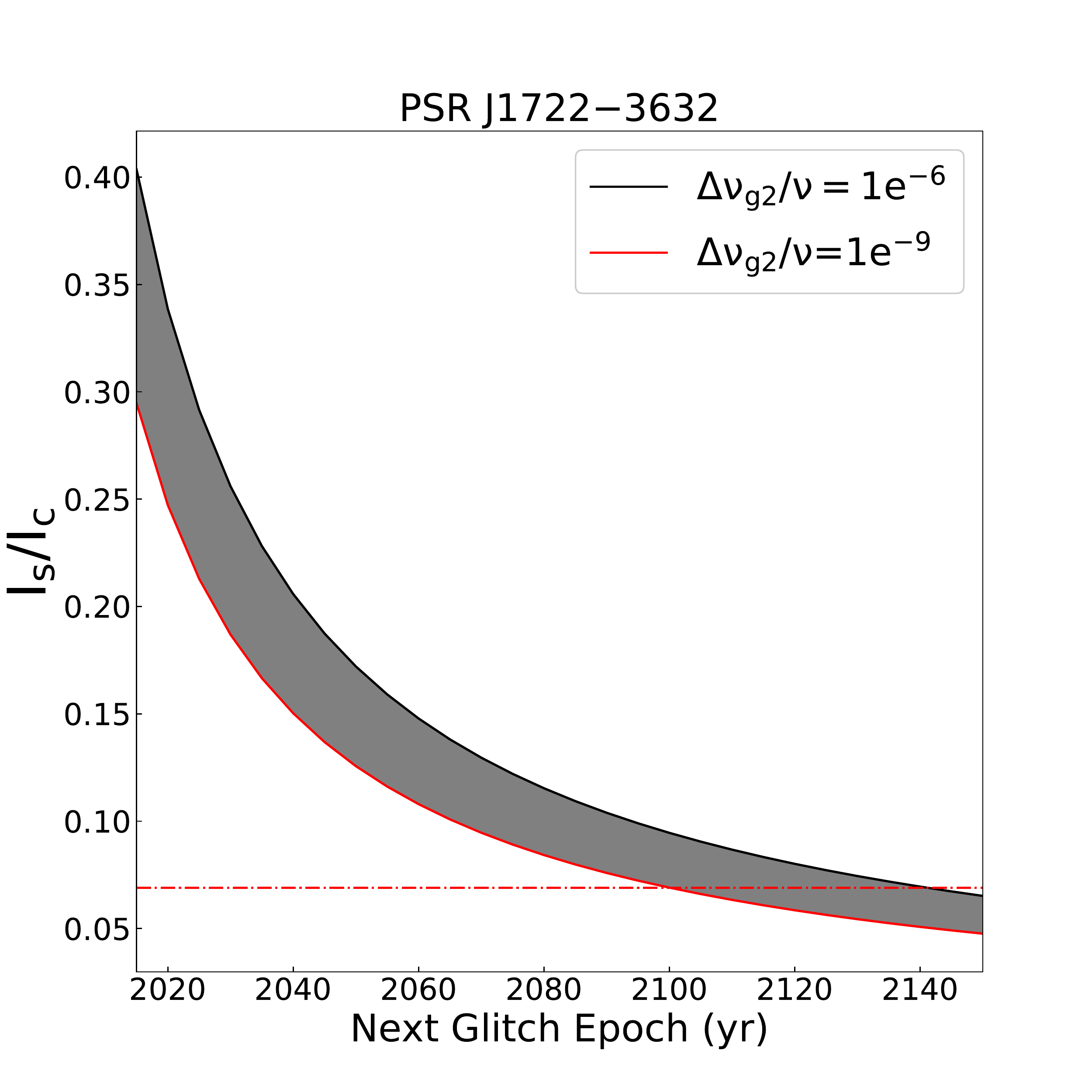}
\includegraphics[width=3.5in,height=3in,angle=0]{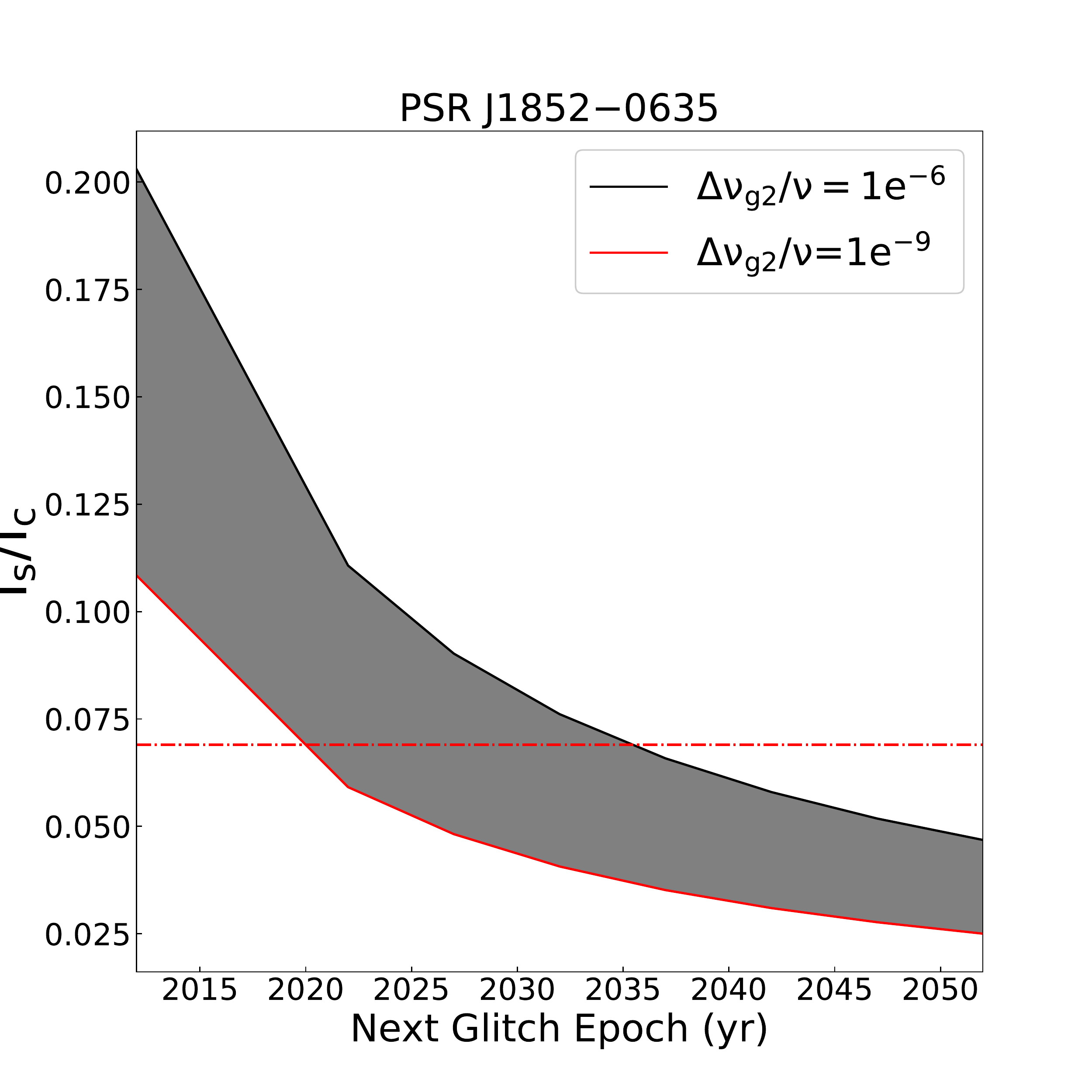}
\caption{Simulated variations of $I_{s}/I_{c}$ for PSRs J1722$-$3632 and J1852$-$0635 in the event of a future glitch. The grey bands outline the values of $I_{s}/I_{c}$ that would result from a glitch in the range $\Delta \nu_g/\nu$ between $10^{-9}$ and $10^{-6}$ at a given epoch. The red dashed line in each panel indicates the value of $\rm I_{s}/I_{c} \sim 6.9\%$ for the Vela pulsar after invoking entrainment.}
\label{fg:sim_Is_Ic}
\end{center}
\end{figure*}

For PSRs J1722$-$3632, J1852$-$0635 and J1957+2831, we find values of $\mathcal{G}$ of about 0.35, 0.11 and 0.0015, respectively. For PSR J1957+2831, $\mathcal{G}$ is less than typical values for young and middle-aged pulsars \citep{heaa15} and can easily be accounted for by angular momentum transfers from just the crustal superfluid. For PSRs J1722$-$3632 and J1852$-$0635, $A_g$ and hence $\mathcal{G}$ are just upper limits. The limits for $\mathcal{G}$ are much larger than the limit for the crustal superfluid, even without invoking entrainment. However, since they are just upper limits, at present these results are not inconsistent with the idea that the crustal superfluid can account for the observed glitches. Detection of another glitch in PSRs J1722$-$3632 or J1852$-$0635 within a few decades would confirm that $\mathcal{G}$ exceeds the crustal superfluid limit. Figure~\ref{fg:sim_Is_Ic} shows the results of simulating the effect of a future glitch for PSRs J1722$-$3632 and J1852$-$0635 on the derived value of $I_{s}/I_{c}$. For PSR J1722$-$3632 in particular, even a small glitch in the next seven decades would result in $I_{s}/I_{c}$ exceeding the crustal superfluid limit even with entrainment. Without entrainment, the intervals would be much longer.

\subsection{Braking Indices}
Under the assumption of pure magneto-dipole radiation, the pulsar braking index $n=3$. However, except in the youngest pulsars, $\ddot{\nu}$ is generally dominated by red-noise processes. \citet{hlk+10} give $\ddot\nu$ for a sample of 366 pulsars. They reported that, for pulsars with characteristic ages $\tau_c > 10^5$~yr, 53\% have a positive $\ddot\nu$ value and the remaining 47\% have negative $\ddot{\nu}$. Also, for a given pulsar, different sections of the data span can have different values of $\ddot\nu$, often changing from positive to negative values. \citet{lkbs19} investigated the detectability of $\ddot\nu$ in the presence of red noise for a sample of millisecond pulsars. Using simulations, it was found that a non-zero $\ddot\nu$ could be recovered when data spans are sufficiently long. \citet{jg99} and \citet{wmz+01} showed that there is a bias toward large positive $\ddot\nu$ in relatively young pulsars and suggested that this could result from the post-glitch decay from unseen earlier glitches.

\begin{figure}
\begin{center}
\centering
\includegraphics[width=3.5in,height=3in,angle=0]{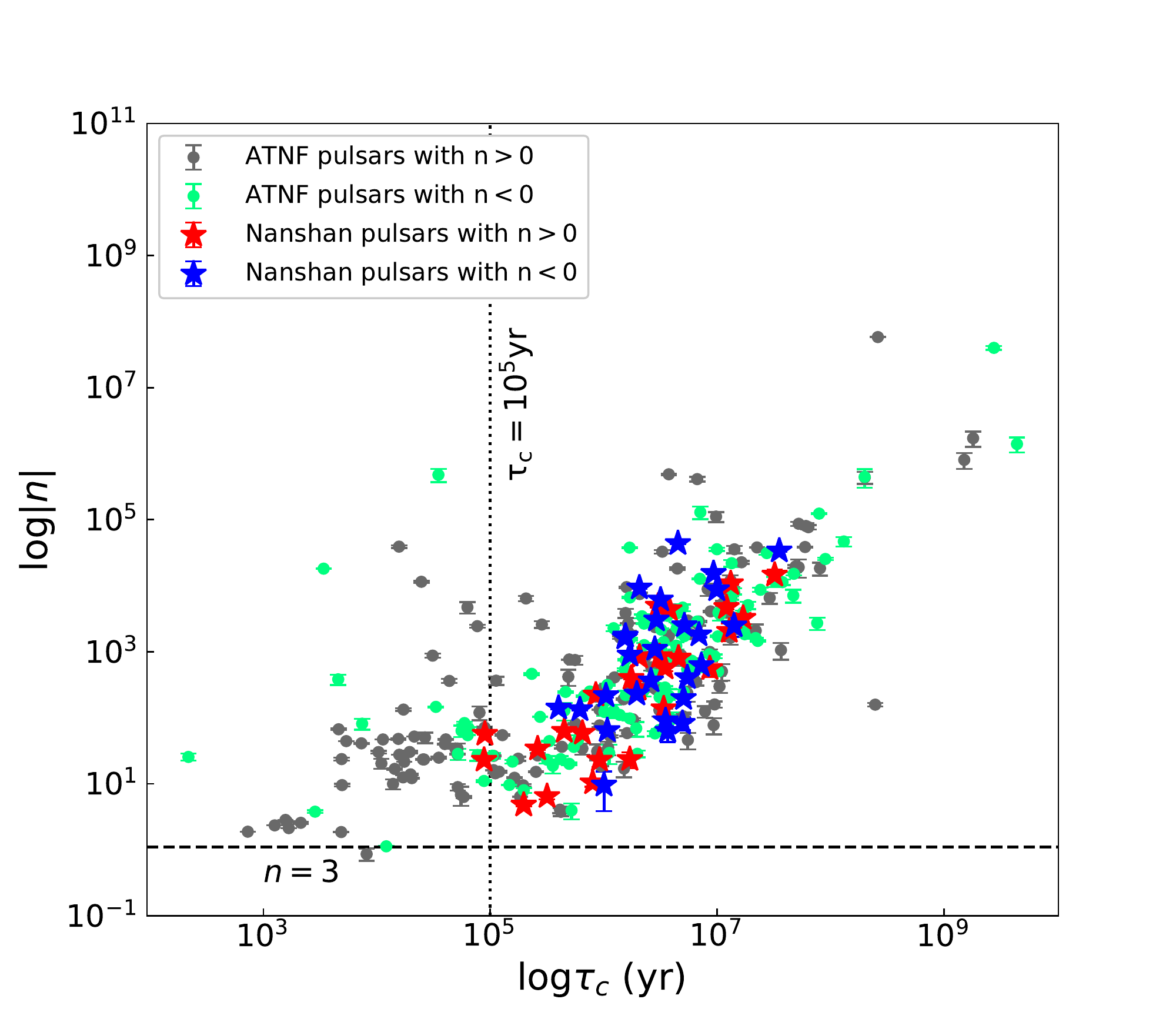}
\caption{ Absolute value of braking index $|n|$ with a significance $>3\sigma$ versus characteristic age $\tau_c$  Gray and green "$\bullet$" stand for the positive and negative $n$ obtained from the ATNF Pulsar Catalogue V1.61 respectively. Red and blue "$\star$" stand for the positive and negative value of $n$ obtained from our sample. The vertical dotted line is at $\tau_c =10^5$~yr and the horizontal dashed line shows $n=3$.}
\label{fg:braking_indices_age}
\end{center}
\end{figure}

In our sample of 87 pulsars, values of frequency second derivative were obtained for 76 pulsars. Excepting PSRs J0614+2229 and J1918+1444 which have $\tau_c \sim 10^4$~yr, most of the sample have $10^5 \la \tau_c \la 10^8$~yr. The range of the value of $\ddot\nu$ is from $-9.9\times 10^{-25}$~s$^{-2}$ to $+54\times 10^{-25}$~s$^{-2}$, and 54\% of the sample have positive values of $\ddot\nu$.  We have also calculated the braking indices for all pulsars in our sample, with a maximum value of $(+3.6\pm 6.3)\times 10^5$ for PSR J1047$-$3032 and a minimum value of $-9.9\pm 3.5)\times 10^5$ for PSR J0520$-$2553. The median value is about 23. These results are consistent with previous results from the literature \citep[e.g.,][]{hlk+04}. Figure~\ref{fg:braking_indices_age} shows the dependence of $|n|$ on $\tau_c$. The data were obtained from the ATNF Pulsar Catalogue V1.61 and our sample; only values with a significance $>3\sigma$ are plotted. It can be seen that for the young pulsars ($\tau_c < 10^5$~yr), no correlation is found between $|n|$ and $\tau_c$ and almost all values of $n$ are positive. For middle-aged and old pulsars ($\tau_c \ga 10^5$~yr), the numbers of positive and negative $|n|$ are almost equal and $|n|$ is strongly correlated with $\tau_c$, with a Pearson correlation coefficient $\rho \approx 0.79$. These results add further support to the idea that observed values of $\ddot\nu$ and $n$ for young pulsars are dominated by glitch recovery, whereas for middle age and older pulsars the anomalous braking indices result from the presence of red timing noise \citep{hlk+10}.

There are many possible sources of such red noise. A likely mechanism is fluctuations in the rate of angular momentum transfer from the interior superfluid to the neutron-star crust, perhaps resulting from superfluid turbulence \citep{ml+14}. Fluctuations in the pulsar magnetosphere can also cause timing irregularities, sometimes related to mode changing \citep[e.g.,][]{klo+06,lhk+10,otkd16}.  \citet{yz15} and \citet{gzy+16} showed that rapid changes in the magnetic inclination angle, including oscillations in the direction of change, could result in red timing noise.

\section{Conclusions}\label{sec:Conclusion}
We have presented a timing analysis of 87 pulsars based on 12 years of observations using the Nanshan radio telescope at Xinjiang Astronomical Observatory. The main conclusions are as follows:

(i) We have obtained positions and proper motions for 48 pulsars. Of these positions, 24 are improved over previously published values. Improved proper motions are given for 21 pulsars and the first published proper motions are given for nine pulsars (Section~\ref{sec:Positions and proper motions}).

(ii) We have obtained transverse velocities for 48 pulsars. Although uncertainties are often large, their statistics are consistent with previous work (Section~\ref{sec:Velocities}).

(iii) We have obtained the rotational parameters for these pulsars. For 36 pulsars, our results have smaller uncertainties than previously published values (Section~\ref{sec:Rotational_parameters}).

(iv) We detected glitches in three pulsars. For two of these (PSRs J1722$-$3632 and J1852$-$0635) they are the first detected glitches. PSR J1722$-$3632 is the second oldest pulsar ($\tau_c \sim 1.4\times 10^6$~yr) in which a large glitch ($\Delta\nu_g/\nu > 10^{-6}$) has been detected (Section~\ref{sec:glitch}).

More frequent timing observations and longer data spans will give more precise rotational parameters, positions and proper motions. Detection of more large glitches, especially for pulsars like PSRs J1722$-$3632 and J1852$-$0635, will further constrain the physics of neutron star interiors.

\section*{Acknowledgements}
We thank the members of the Pulsar Group at XAO, Lunhua Shang and Weihua Wang for useful discussions. This work is supported by the National Key Research and Development Program of China (2016YFA0400804), the Operation, Maintenance and Upgrading Fund for Astronomical Telescopes and Facility Instruments, budgeted from the Ministry of Finance of China (MOF) and administrated by the Chinese Academy of Sciences (CAS), National Natural Science Foundation of China (No.11873080). Lin Li is supported by National Natural Science Foundation of China (No.11463005) and Doctoral Research Star-up Fund at Xinjiang University. Wang Jingbo is supported by the Youth Innovation Promotion Association of Sciences, $201^{*}$ Project of Xinjiang Uygur Autonomous Region of China for Flexibly Fetching in Upscale Talents. Liu Zhiyong is supported by the National Key R $\&$D Program of China under grant number 2018YFA0404603.





\end{document}